\documentclass[preprint]{elsarticle}

\usepackage{epsfig}
\usepackage{graphicx}
\usepackage{amssymb}
\usepackage{algorithm}
\usepackage{moreverb}
\usepackage{multirow}
\usepackage{rotating}
\usepackage{subfigure}

\usepackage[noend]{algcompatible}

\algblockdefx{FORALLP}{ENDFAP}[1]%
  {\textbf{for all }#1 \textbf{ in parallel}}

\begin{document}
\title{Efficient Irregular Wavefront Propagation Algorithms on Hybrid CPU-GPU Machines}
\author{George Teodoro\corref{fn1}}
\ead{george.teodoro@emory.edu}
\author{Tony Pan}
\ead{tony.pan@emory.edu}
\author{Tahsin Kurc}
\ead{tkurc@emory.edu} 
\author{Jun Kong}
\ead{jun.kong@emory.edu}
\author{Lee Cooper}
\ead{lee.cooper@emory.edu}
\author{Joel Saltz}
\ead{jhsaltz@emory.edu}
\address{Center for Comprehensive Informatics and Biomedical Informatics Department, \\
Emory University, Atlanta, GA 30322}

\cortext[fn1]{Corresponding author}
\begin{abstract}
In this paper, we address the problem of efficient execution of a computation
pattern, referred to here as the irregular wavefront propagation pattern
(IWPP), on hybrid systems with multiple CPUs and GPUs. The IWPP is common in
several image processing operations. In the IWPP, data elements in the
wavefront propagate waves to their neighboring elements on a grid if a
propagation condition is satisfied. Elements receiving the propagated waves
become part of the wavefront. This pattern results in irregular data accesses
and computations.  We develop and evaluate strategies for efficient computation
and propagation of wavefronts using a multi-level queue structure. This queue
structure improves the utilization of fast memories in a GPU and reduces
synchronization overheads. We also develop a tile-based parallelization
strategy to support execution on multiple CPUs and GPUs.  We evaluate our
approaches on a state-of-the-art GPU accelerated machine (equipped with 3 GPUs
and 2 multicore CPUs) using the IWPP implementations of two widely used image
processing operations: morphological reconstruction and euclidean distance
transform. Our results show significant performance improvements on GPUs. The
use of multiple CPUs and GPUs cooperatively attains speedups of 50$\times$ and
85$\times$ with respect to single core CPU executions for morphological
reconstruction and euclidean distance transform, respectively.


\end{abstract}


\maketitle

\section{Introduction} \label{sec:intro}
This paper investigates efficient parallelization on hybrid CPU-GPU systems of
operations or applications whose computation structure includes what we call
the {\em irregular wavefront propagation pattern} (IWPP) (see
Algorithm~\ref{alg:genericRecon}). Our work is motivated by the requirements of
analysis of whole slide tissue images in biomedical research. With rapid
improvements in sensor technologies and scanner instruments, it is becoming
feasible for research projects and healthcare organizations to gather large
volumes of microscopy images.  We are interested in enabling more effective use
of large datasets of high resolution tissue slide images in research and
patient care.  A typical image from state-of-the-art scanners is about
50K$\times$50K to 100K$\times$100K pixels in resolution.  A whole slide tissue
image is analyzed through a cascade of image normalization, object
segmentation, object feature computation, and object/image classification
stages. The segmentation stage is expensive and composed of a pipeline of
substages.  The most expensive substages are built on several low-level
operations, notably morphological
reconstruction~\cite{Vincent93morphologicalgrayscale} and distance
transform~\cite{Vincent:1991:ExEuDi}. Efficient implementation of these
operations is necessary to reduce the cost of image analysis. 

Processing a high resolution image on a single CPU system can take hours.  The
processing power and memory capacity of graphics processing units (GPUs) have
rapidly and significantly improved in recent years. Contemporary GPUs provide
extremely fast memories and massive multi-processing capabilities, exceeding
those of multi-core CPUs. The application and performance benefits of GPUs for
general purpose processing have been demonstrated for a wide range of
applications~\cite{Fialka:2006:FCP:1153927.1154738,Scholl:2011:CMI:1938207.1938216,icl523,1587427,1015800,ravi2010compiler}.
As a result, CPU-GPU equipped machines are emerging as viable high performance
computing platforms for scientific computation~\cite{10.1109/MCSE.2011.83}. 

The processing structures of the morphological reconstruction and distance
transform operations bear similarities and include the  IWPP on a grid. The
IWPP is characterized by one or more source grid points from which waves
originate and the irregular shape and expansion of the wave fronts. The
composition of the waves is dynamic, data dependent, and computed during
execution as the waves are expanded. Elements in the front of the waves work as the
sources of wave propagations to neighbor elements. A propagation occurs only when
a given {\em propagation condition}, determined based on the value of a
wavefront element and the values of its neighbors, is satisfied. In practice,
each element in the propagation front represents an independent wave
propagation; interaction between waves may even change the direction of the
propagation. In the IWPP only those elements in the wavefront are the ones
effectively contributing for the output results. Because of this property, an
efficient implementation of irregular wavefront propagation can be accomplished
using an auxiliary container structure, e.g., a queue, set, or stack, to keep
track of active elements forming the wavefront. The basic components of the
IWPP are shown in Algorithm~\ref{alg:genericRecon}. 

\begin{algorithm}
\caption{Irregular Wavefront Propagation Pattern (IWPP)}
\label{alg:genericRecon}
\begin{algorithmic}[1]
\STATE $D \leftarrow$ data elements in a multi-dimensional space
\STATE \{{\bf Initialization Phase}\}
\STATE $S \leftarrow$ subset active elements from $D$
\STATE \{{\bf Wavefront Propagation Phase}\}
\WHILE{$S \neq \emptyset$}
	\STATE Extract $e_i$ from $S$
	\STATE $Q \leftarrow$ $N_G(e_i)$
	\WHILE{$Q \neq \emptyset$}
		\STATE Extract $e_j$ from $Q$
		\IF{$PropagationCondition$($D(e_i)$,$D(e_j)$) $=$ true}
			\STATE $D(e_j) \leftarrow$ $Update$($D(e_i)$)
			\STATE Insert $e_j$ into $S$
		\ENDIF
	\ENDWHILE
\ENDWHILE 
\end{algorithmic}
\end{algorithm}

In this algorithm, a set of elements in a multi-dimensional grid space ($D$)
are selected to form the initial wavefront ($S$). These active elements then
act as wave propagation sources in the wavefront propagation phase.  During
propagation phase, a single element ($e_i)$ is extracted from the wavefront and
its neighbors ($Q \leftarrow N_G(e_i)$) are identified.  The neighborhood of an
element $e_i$ is defined by a discrete grid $G$, also referred to as the
structuring element. The element $e_i$ tries to propagate the wavefront to each
neighbor $e_j \in Q$. If the propagation condition ($PropagationCondition$),
based on the values of $e_i$ and $e_j$, is satisfied, the value of the element
$e_j$ ($D(e_j)$) is updated, and $e_j$ is inserted in the container ($S$). The
assignment operation performed in Line 11, as a consequence of the wave
expansion, is expected to be commutative and atomic. That is, the order in
which elements in the wavefront are computed should not impact the
algorithm results. The wavefront propagation process continues until 
stability is reached; i.e., until the wavefront container is empty.

The IWPP is not unique to morphological reconstruction and distance transform. Core 
computations in several other image processing methods contain a similar structure: 
Watershed~\cite{citeulike:557456}, Euclidean skeletons~\cite{meyer90digital}, 
and skeletons by influence zones~\cite{lantuejoul80}. Additionally, Delaunay
triangulations~\cite{Preparata:1985:CGI:4333}, Gabriel
graphs~\cite{citeulike:3982757} and relative neighborhood
graphs~\cite{Toussaint_1980} can be derived from these methods. 
Another example of the IWPP 
is the shortest-path computations on a grid which contains obstacles 
between source points and destination points (e.g., shortest path calculations in
integrated circuit designs). A high performance implementation of the IWPP can 
benefit these methods and applications.  

The traditional wavefront computation is common in many scientific
applications~\cite{springerlink:10.1007/BF01407876,banerjee91double,Ancourt:1991:SPL:109625.109631}.
It is also a well-known parallelization pattern in high performance computing.
In the classic wavefront pattern, data elements are laid out in a
multi-dimensional grid. The computation of an element on the wavefront depends
on the computation of a set of neighbor points.  The classic wavefront pattern
has a regular data access and computation structure in that the wavefront
starts from a corner point of the grid and sweeps the grid diagonally from one
region to another.  The morphological reconstruction and distance transform
operations could potentially be implemented in the form of iterative traditional wavefront
computations. However, the IWPP offers a more efficient execution structure,
since it avoids touching and computing on data points that do not contribute to
the output. The IWPP implementation on a hybrid
machine consisting of multiple CPUs and GPUs is a challenging problem. The difficulties with IWPP
parallelization are accentuated by the irregularity of the computation that is
spread across the input domain and evolves during the computation as active
elements change. 

In this paper, we propose, implement,
and evaluate parallelization strategies  
for efficient execution of IWPP computations on large grids (e.g., 
high resolution images) on a machine with multiple CPUs and GPUs. 
The proposed strategies can
take full advantage this computation pattern characteristics to achieve
efficient execution on multicore CPU and GPU systems. 
The contributions of our work can be summarized as follows: 
\begin{itemize}
\item We identify a computation pattern commonly found in several analysis 
operations: the irregular wavefront propagation pattern. 
\item We develop an efficient GPU implementation of the IWPP algorithm using a multi-level 
queue structure. The multi-level queue structure improves the utilization of fast memories 
in a GPU and reduces synchronization overheads among GPU threads.
To the best of our knowledge this is the first work to implement this
pattern for GPU accelerated environments. 
\item We develop efficient implementations of the morphological reconstruction
and distance transform operations on a GPU using the GPU-enabled IWPP
algorithm. These are the first GPU-enabled IWPP-based implementations of
morphological reconstruction and distance transform. The morphological
reconstruction implementation achieves much better performance than a previous
implementation based on raster/anti-raster scans of the image.  The output
results for morphological reconstruction and distance transform are exact
regarding their sequential counterparts.
\item We extend the GPU implementation of the IWPP to support processing of images that 
do not fit in GPU memory through coordinated use of multiple CPU cores and multiple GPUs 
on the machine. 
\end{itemize}
We perform a performance evaluation of the IWPP implementations using the
morphological reconstruction and distance transform operations on a
state-of-the-art hybrid machine with 12 CPU cores and 3 NVIDIA Tesla GPUs.
Significant performance improvements were observed in our experiments. Speedups
of 50$\times$ and 85$\times$ were achieved, as compared to the single CPU core
execution, respectively, for morphological reconstruction and euclidean
distance transform when all the CPU cores and GPUs were used in a coordinated
manner. We were able to compute morphological reconstruction on a
96K$\times$96K-pixel image in 21 seconds and distance transform on a
64K$\times$64K-pixel image in 4.1 seconds.  These performances make it feasible
to analyze very high resolution images rapidly and conduct large scale studies.   

The manuscript is organized as follows. Section~\ref{sec:wavefront} presents
the use of the IWPP on image analysis, including preliminary definitions and
the implementation of use case algorithms: morphological reconstruction and
euclidean distance transform. The IWPP parallelization strategy and its support
in GPUs are discussed in Section~\ref{sec:wavegpu}. The extensions for multiple
GPUs and cooperative CPU-GPU execution are described in
Section~\ref{sec:multigpu}. The experimental evaluation is presented in
Section~\ref{sec:results}. Finally, Sections~\ref{sec:related}
and~\ref{sec:conc}, respectively, present the related work and conclude the
paper.

\section{Irregular Wavefront Propagation Pattern in Image Analysis} \label{sec:wavefront}
We describe two morphological algorithms, {\em Morphological Reconstruction}
and {\em Euclidean Distance Transform}, from the image analysis domain to illustrate the
IWPP.  Morphological operators are basic operations used by a broad set of
image processing algorithms. These operators are applied to individual pixels
and are computed based on the current value of a pixel and pixels in its
neighborhood.  A pixel $p$ is a neighbor of pixel $q$ if $(p,q) \in G$. $G$ is
usually a 4-connected or 8-connected square grid.  $N_G(p)$ refers to the set
of pixels that are neighbors of $p \in \mathbb{Z}^n$ according to $G$ ($N_G(p)
= \{q \in \mathbb{Z}^n| (p,q) \in G\}$).  Input and output images are defined
in a rectangular domain $D_I \in \mathbb{Z}^n \rightarrow \mathbb{Z}$.  The
value $I(p)$ of each image pixel $p$ assumes 0 or 1 for binary images. For gray
scale images, the value of a pixel comes from a set $\{0, ... , L-1\}$ of gray
levels from a discrete or continuous domain.  

\subsection{Morphological Reconstruction Algorithms} \label{mr-algo} 

\begin{figure}[ht]
\begin{center}
\includegraphics[width=0.9\textwidth]{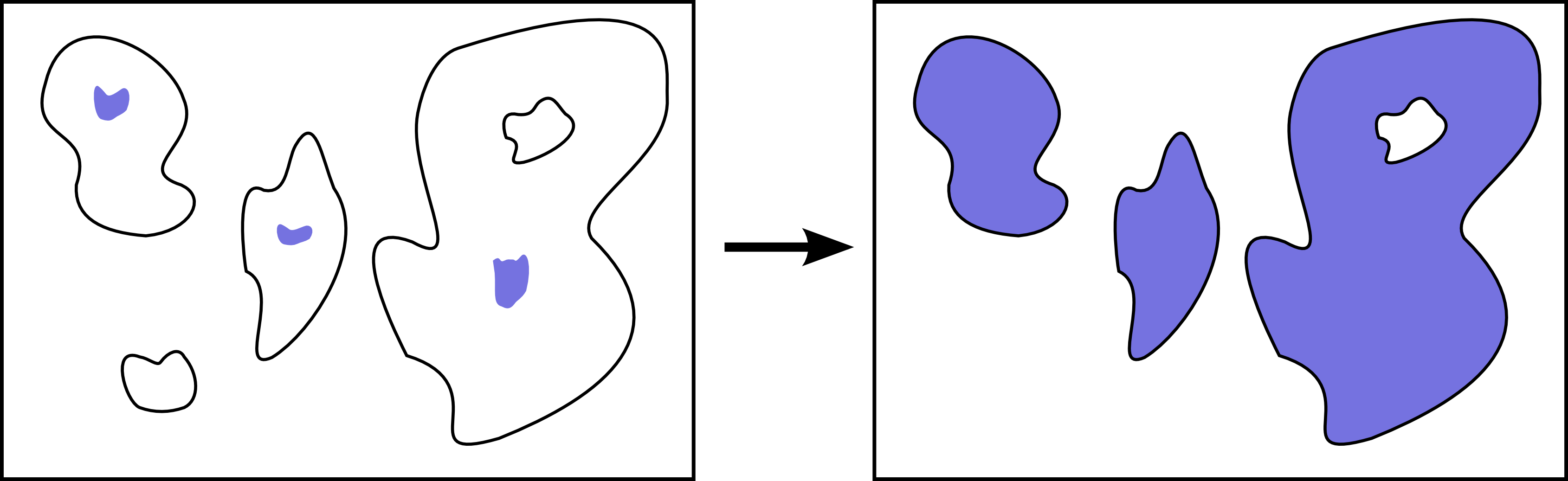}
\vspace*{-2ex}
\caption{Binary morphological reconstruction from markers. The markers are the dark 
patches inside three of the objects in the image on the left. The image on the 
right shows the reconstructed objects and the final binary image after the application 
of morphological reconstruction.}
\vspace*{-2ex}
\label{fig:morph-reconst}
\end{center}
\end{figure}

Morphological reconstruction is one of the elementary operations in image 
segmentation~\cite{Vincent93morphologicalgrayscale}. When applied to binary 
images, it pulls out the connected components of 
an image identified by a {\em marker} image. Figure~\ref{fig:morph-reconst} 
illustrates the process. 
The dark patches inside three objects in image 
$I$ (also called the mask image) on the left correspond to the marker image 
$J$. The result of the morphological reconstruction is shown on the right.

\begin{figure}[ht]
\begin{center}
\includegraphics[width=0.9\textwidth]{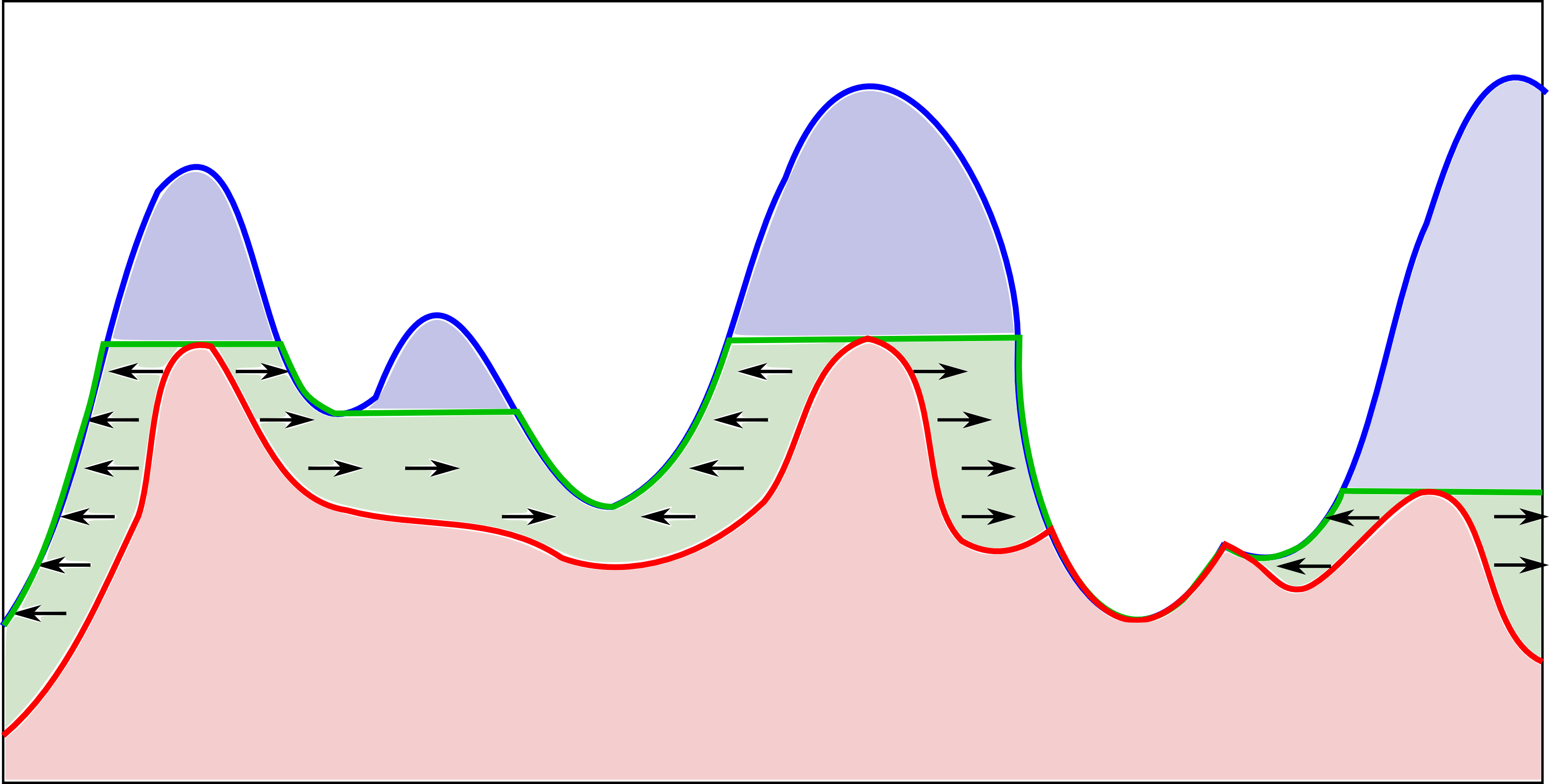}
\vspace*{-2ex}
\caption{Gray scale morphological reconstruction in 1-dimension. The marker image 
intensity profile is represented as the red line, and the mask image intensity profile 
is represented as the blue line.  The final image intensity profile is represented as 
the green line. The arrows show the direction of propagation from the marker intensity 
profile to the mask intensity profile.  The green region shows the changes introduced 
by the morphological reconstruction process.}
\vspace*{-2ex}
\label{fig:grayscale-morph-reconst}
\end{center}
\end{figure}

Morphological reconstruction can also be applied to gray scale images. 
Figure~\ref{fig:grayscale-morph-reconst} illustrates the process of gray scale
morphological reconstruction in 1-dimension.  The marker intensity profile is
propagated spatially but is bounded by the mask image's intensity profile. The
primary difference between binary and gray scale morphological reconstruction
algorithms is that in binary reconstruction, any pixel value change is
necessarily the final value change, whereas a value update in gray scale
reconstruction may later be replaced by another value update.  In color
images, morphological reconstruction can be applied either to individual
channels (e.g., the Red, Green, and Blue channels in an RGB image) or to gray
scale values computed by combining the channels.

The morphological reconstruction $\rho_{I}(J)$ of
mask $I$ from marker image $J$ is done by performing elementary dilations
(i.e., dilations of size 1) in $J$ by $G$, the structuring element. 
An elementary dilation from a pixel
$p$ corresponds to propagation from $p$ to its immediate neighbors in $G$. The
basic algorithm carries out elementary dilations successively over the entire
image $J$, updates each pixel in $J$ with the pixelwise minimum of the
dilation's result and the corresponding pixel in $I$ (i.e., $J(p) \leftarrow
(max\{J(q), q \in N_G(p) \cup \{p\}\}) \wedge I(p)$, where $\wedge$ is the 
pixelwise minimum operator), and stops when {\em stability} is reached, i.e., 
when no more pixel values are modified. Several morphological reconstruction 
algorithms for gray scale images have been developed by 
Vincent~\cite{Vincent93morphologicalgrayscale} based on this core technique. We 
present brief descriptions of these algorithms below and refer the reader to the 
original paper for more details. \\

\noindent {\em Sequential Reconstruction (SR)}: Pixel value propagation in the
marker image is computed by alternating raster and anti-raster scans. A raster 
scan starts from the pixel at $(0,0)$ and proceeds to the pixel at $(N-1,M-1)$ in 
a row-wise manner, while an anti-raster starts from the pixel at $(N-1,M-1)$ and 
moves to the pixel at $(0,0)$ in a row-wise manner. Here, $N$ and $M$ are the resolution 
of the image in x and y dimensions, respectively.  In each scan, values from pixels in the upper 
left or the lower right half neighborhood are propagated to the current pixel in raster 
or anti-raster fashion, respectively.
The raster and anti-raster scans  
allow for changes in a pixel to be propagated in the current
iteration. The SR method iterates until stability is reached, i.e., no more changes in pixels 
are computed. \\

\noindent {\em Queue-based
Reconstruction (QB)}: In this method, a first-in first-out (FIFO)
queue is initialized with pixels in the regional maxima. The
computation then proceeds by removing a pixel from the queue, scanning 
the pixel's neighborhood, and queuing the neighbor pixels whose values 
have been changed. The overall process continues until the queue
is empty. \\ 

\noindent {\em Fast Hybrid Reconstruction (FH)}: 
The computation of the regional maxima needed to initialize the queue 
in QB incurs significant computational cost. The FH approach incorporates
the characteristics of the SR and QB algorithms to reduce the cost of initialization, 
and is about one order of
magnitude faster than the others. It first makes one pass using the raster and
anti-raster orders as in SR. After that pass, it continues the computation
using a FIFO queue as in QB. A pseudo-code implementation of FH is presented in
Algorithm~\ref{alg:fh}, $N^+_G$ and $N^-_G$ denote the set of neighbors in 
$N_G(p)$ that are reached before and after 
touching pixel $p$ during a raster scan. 
\begin{algorithm}
\caption{Fast Hybrid Gray scale Morphological Reconstruction Algorithm}
\label{alg:fh}
{\bf Input}
\vspace{-2.5mm}
\begin{description}
	\item[$I$:]      \emph{mask image}
	\vspace{-2.5mm}
	\item[$J$:]      \emph{marker image.}
\end{description}
\vspace{-2.5mm}
\begin{algorithmic}[1]

	\STATE \{{\bf Initialization Phase}\}
	\STATE Scan $I$ and $J$ in raster order.
		\STATE{\ \ Let $p$ be the current pixel}
%
		\STATE{\ \ $J(p) \leftarrow (max\{J(q), q \in N_G^+(p) \cup \{p\}\}) \wedge I(p)$}
	\STATE Scan $I$ and $J$ in anti-raster order.
		\STATE{\ \ Let $p$ be the current pixel}
		\STATE{\ \ $J(p) \leftarrow (max\{J(q), q \in N_G^-(p) \cup \{p\}\}) \wedge I(p)$}
		\STATE{\ \ {\bf if\ }{$\exists q\in N_G^-(p)\ |\ J(q)<J(p)\ and\ J(q)<I(q)$}}
			\STATE{\ \ \ \ queue\_add(p)}
\STATE \{{\bf Wavefront Propagation Phase}\}
\WHILE{queue\_empty() = false}
	\STATE $p \leftarrow$ dequeue()
	\FORALL{$q \in N_{G}(p)$}
		\IF{$J(q) < J(p)$ and $I(q) \neq J(q)$}
			\STATE $J(q) \leftarrow min\{J(p),I(q)\}$
			\STATE queue\_add(q)
		\ENDIF
	\ENDFOR
\ENDWHILE
\end{algorithmic}
\end{algorithm}
\subsection{Euclidean Distance Transform} \label{sec:dt_algo}
The distances transform (DT) operation computes a distance map $M$ from a
binary input image $I$, where for each pixel $p \in I$ the pixel's value
in $M$, $M(p)$, is the smallest distance from $p$ to a background pixel. 
DT is a fundamental operator in shape analysis, which can be used for  
separation of overlapping objects in watershed based 
segmentation~\cite{citeulike:557456,insilico}. It can also be used in the 
calculation of morphological operations~\cite{Cuisenaire1999163} as erosion and dilation and 
the computation of Voronoi diagrams and Delaunay triangulation~\cite{Vincent:1991:ExEuDi}.

The definition of DT is simple, but historically it has been 
hard to achieve good precision and efficiency for this
operation. For a comprehensive discussion of algorithms for calculation of distance transform,
we refer the reader to the survey by Fabbri et al.~\cite{Fabbri082deuclidean}. 
The first DT algorithms were proposed by Rosenfeld and
Pfaltz~\cite{Rosenfeld:1966:SOD:321356.321357}. These algorithms were based on
raster/anti-raster scans strategies to calculate non-Euclidean metrics such as cityblock and
chessboard, which at that time were used as approximations to the Euclidean
distance. Danielsson proposed an algorithm~\cite{danielsson80} to compute Euclidean DT
that propagates information of the nearest background pixel using neighborhood
operators in a raster/anti-raster scan manner. With this strategy, a two-element vector
with location information of the nearest background pixel is propagated
through neighbor pixels. Intrinsically, it builds Voronoi diagrams where
regions are formed by pixels with the same nearest background pixel. 

\begin{figure}[ht]
\begin{center}
\includegraphics[width=0.45\textwidth]{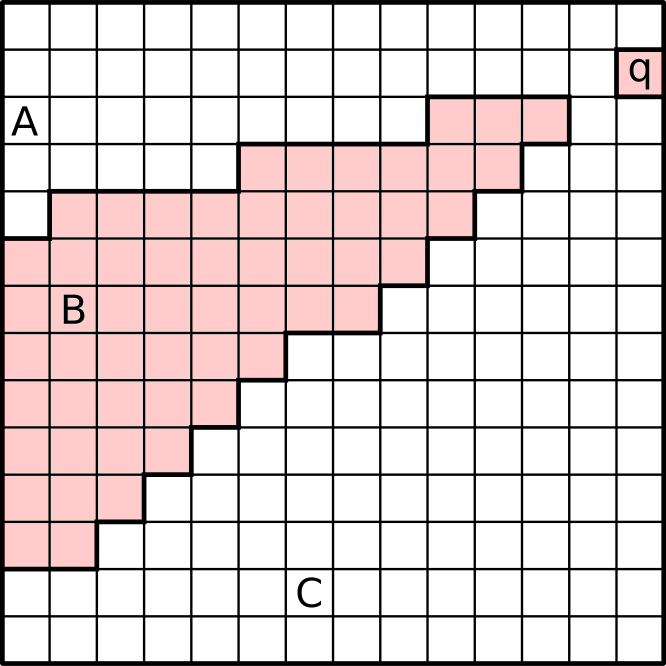}
\vspace*{-2ex}
\caption{Errors introduced by neighborhood propagation in the euclidean distance 
transform operation using an 8-connected neighborhood method (Cuisenaire and Macq~\cite{Cuisenaire1999163}). 
Pixel $q$ is closer to object B than it is to A and C. However, $q$ is not neighbor to any pixel in 
the connected Voronoi diagram starting at B. The distance value assigned to $q$, $M(q)$, would 
be $\ \sqrt[]{170}$ (about 13.038), instead of $ \sqrt[]{169}$ (13) in the exact case.}
\vspace*{-2ex}
\label{fig:voronoi}
\end{center}
\end{figure}

Danielsson's algorithm is not an exact algorithm. The Voronoi
diagram is not connected in a discrete space, though it is in the continuous
space. Hence, the Voronoi diagram computation through a neighborhood
based operator introduces approximation errors as illustrated in
Figure~\ref{fig:voronoi}.  Nevertheless, approximation errors introduced by the
algorithm are small and bound by a mathematical framework. Approximations
computed by this algorithm have been found to be sufficient in practice for
several
applications~\cite{daSTorres20041163,DBLP:journals/pr/FalcaoCC02,1028280}.
Moreover, some of the exact euclidean distance transform
algorithms~\cite{Cuisenaire1999163,Shih_Wu_2004,Mullikin1992526} are
implemented as a post-processing phase to Danielsson's algorithm, in which
approximation errors are resolved. 

Algorithm~\ref{alg:dt} presents an irregular wavefront propagation based approach to
efficiently compute the single neighborhood approximation of DT. In the
\emph{initialization phase} a pass on the input is performed to identify and
queue pixels forming the initial wavefront.  In Lines~2~and~3, Voronoi diagram
(VR) that holds the current nearest background pixel for each image pixel is
initialized. Each background pixel will itself be its nearest pixel, while the
foreground pixels will initially point to a pixel virtually at infinite
distance ($inf$).  The $inf$ pixel is such that for each pixel $p$, $p$ is
closer to any other pixel in the image domain than it is to the $inf$ pixel. At
the end of the initialization phase, background pixels with a foreground
neighbor (contour pixels) are added to a queue for propagation
(Lines~4~and~5). 

\begin{algorithm}
\caption{Wavefront Propagation-based Distance Transform Algorithm}
\label{alg:dt}
\begin{description}
	\item[{\bf Input:}$I$:]      \emph{mask image}
\vspace{-2.5mm}
	\item[{\bf Output:}$M$:]      \emph{distance map image}
\vspace{-2.5mm}
\end{description}
\begin{description}
	\item[$FG$:] \emph{Foreground}; $BG$: \emph{Background}
\end{description}
\vspace{-2.5mm}
\begin{algorithmic}[1]
\STATE \{{\bf Initialization Phase}\}
	\FORALL {$p \in D_I$}
		\STATE{$VR(p) = ( p==BG)\ ?\ p\ :\ inf$}
		\IF{$I(p) == BG\ and\ \exists q\in N_G(p)\ |\ I(q) == FG$}
			\STATE{\ \ \ queue\_add(p)}
		\ENDIF
	\ENDFOR
\STATE \{{\bf Wavefront Propagation Phase}\}
\WHILE{queue\_empty() = false}
	\STATE $p \leftarrow$ dequeue()
	\FORALL{$q \in N_{G}(p)$}
		\IF{DIST$(q, VR(p)) <$ DIST$(q, VR(q))$}
			\STATE $VR(q) = VR(p)$
			\STATE queue\_add(q)
		\ENDIF
	\ENDFOR
\ENDWHILE
\FORALL {$p \in D_I$}
	\STATE $M(p) = $ DIST$(p, VR(p))$)
\ENDFOR
\end{algorithmic}
\end{algorithm}

The \emph{wavefront propagation phase} of the EDT is carried out in Lines 7 to
12. During each iteration, a pixel $p$ is removed
from the queue. The pixel's nearest background pixel is propagated to the
neighbors through a neighborhood operator. For each neighbor pixel $q$, if the distance of 
$q$ to the current assigned nearest background pixel (DIST$(q, VR(q))$) is
greater than the distance of $q$ to the nearest background pixel of $p$ (DIST$(q, VR(p))$),
the propagation occurs. In that case, the nearest background pixel of $q$ is
updated with that of $p$ (Line 11), and $q$ is added to the queue
for propagation. After the propagation phase is completed, the distance map is computed 
from the Voronoi diagram.

As presented in~\cite{Cuisenaire1999163}, using a sufficiently large
neighborhood (structuring element) will guarantee the exactness of this
algorithm.  In addition, the wavefront propagation method could be extended as
shown in~\cite{Vincent:1991:ExEuDi} to compute the exact distance transform. To
include these extensions, however, another queue would have to be used,
increasing the algorithm cost. As discussed before, Danielsson's distance
transform provides acceptable results for a large number of
applications~\cite{daSTorres20041163,DBLP:journals/pr/FalcaoCC02}.  We have
employed this version in our image analysis applications.

\section{Implementation of Irregular Wavefront Propagation Pattern on a GPU}
\label{sec:wavegpu}

We first describe the strategy for parallel execution of the IWPP on GPUs. 
We then present a parallel queue
data structure that is fundamental to efficiently tracking data elements in the
wavefront. In the last two sections, we describe the GPU-enabled
implementations of the two operations presented 
in Section~\ref{sec:wavefront}. 

\subsection{Parallelization Strategy} 

After the initialization phase of the IWPP, active data elements forming the initial
wavefront are put into a global queue for computation. To allow parallel
execution, the data elements in the global queue are partitioned into a number
of smaller queues. Each of these queues is assigned to a GPU thread block. Each
block can carry out computations independent of the other blocks.  This
partitioning and mapping strategy reduces the communication and synchronization
costs to access the global queue. With this
strategy, it is not necessary to employ expensive inter-block
synchronizations~\cite{DBLP:conf/ipps/XiaoF10} or continuously utilize system
level $Memory\ Fence\ Operations$ to assert queue consistency across GPU
multi-processors.

The execution of each neighborhood operator is performed considering the value
of the current data element being  processed ($p$) and that of each data element 
in the neighborhood ($q \in N_G(p)$). To correctly perform this computation in
parallel, it is necessary to guarantee atomicity in updates on neighbor 
data elements ($q$). This can only be achieved at the cost of extra synchronization
overheads. Fortunately, the use of atomic compare-and-swap (CAS) operations 
is sufficient to solve this race condition.  
The efficiency of atomic operations in GPUs have been
significantly improved in the last generation of NVIDIA GPUs
(Fermi)~\cite{citeulike:8927897} because of the use of cache memory. Atomic
operations obviously still are more efficient in cases where threads do not
concurrently try to update the same memory address.  When multiple threads
attempt to update the same memory location, they are serialized in order
to ensure correct execution. As a result, the number of operations successfully
performed per cycle might be drastically reduced~\cite{citeulike:8927897}. To 
lessen the impact of potential serialization, our GPU parallelization employs 
data element level parallelism which allows for data elements queued for computation 
to be independently processed.

\begin{algorithm}
\caption{Wavefront propagation phase on a GPU}
\label{alg:queuephasegpu}
\begin{algorithmic}[1]
\STATE\{Split initial queue equally among thread blocks\}
\WHILE{queue\_empty() = false}
 	\WHILE{$(p = dequeue(...)) !=\ $EMPTY} {\bf in parallel}
		\FORALL{$q \in N_{G}(p)$}
			\REPEAT
				\STATE $curValueQ = I(q)$ 
				\IF{$PropagationCondition(I(p), curValueQ)$}
					\STATE $oldval = atomicCAS(\&I(q), op(I(p)), curValueQ)$
					\IF{$oldval\ != curValueQ$}
						\STATE queue\_add(q)
						\STATE $break$;
					\ENDIF
				\ELSE
					\STATE $break$;
				\ENDIF
			\UNTIL{True}
		\ENDFOR
	\ENDWHILE
		\STATE{queue\_swap\_in\_out()}
\ENDWHILE
\end{algorithmic}
\end{algorithm}

The GPU-based implementation of the propagation operation is presented in
Algorithm~\ref{alg:queuephasegpu}. After splitting the initial queue, each
block of threads enter into a loop in which data elements are dequeued in
parallel and processed, and new data elements may be added to the local queue
as needed.  This process continues until the queue is empty. Within each loop,
the data elements queued in the last iteration are uniformly divided among the
threads in the block and processed in parallel (Lines 3---14 in
Algorithm~\ref{alg:queuephasegpu}). The value of each queued data element $p$
is compared to every data element $q$ in its neighborhood.  An atomic operation
is performed when a neighbor data element $q$ should be updated
($PropagationCondition$ is evaluated true).  The value of data element $q$
before the compare and swap operation is returned ($oldval$) and used to
determine whether its value has really been changed (Line 9 in
Algorithm~\ref{alg:queuephasegpu}). This step is necessary because there is a
chance that another thread might have changed the value of $q$ between the time
the data element's value is read to perform the propagation condition test and
the time the atomic operation is performed (Lines 6--8 in
Algorithm~\ref{alg:queuephasegpu}). If the atomic compare and swap operation
performed by the current thread has changed the value of data element $q$, $q$
is added to the queue for processing in the next iteration of the loop.  Even
with this control, it is possible that between the test in line 9 and the
addition to the queue, the data element $q$ may have been modified again. In
this case, $q$ is added multiple times to the queue. Although it impacts
the performance, the correctness of the algorithm is not affected because the
update operations replace the value of $q$ via a commutative and atomic
assignment operation. 

After computing the data elements from the last iteration, data elements that are
added to the queue in the current iteration are made available for the next
iteration. Elements used as input in one iteration and those
queued during the propagation are stored in different queues.  The
process of making elements queued available includes swapping the input and
output queues (queue\_swap\_in\_out(); Lines~15). As
described in the next section, the choice to process data elements in rounds,
instead of making each data element inserted into the queue immediately
accessible, is made in our design to implement a queue with very efficient read
performance (dequeue performance) and with low synchronization costs to provide
consistency among threads when multiple threads read from and write to the
queue. When maximum and minimum operations are used in the propagation condition 
test, $atomicCAS$ may be replaced with an $atomicMax$ or $atomicMin$ operation 
to improve performance.

Since elements in the queue are divided among threads for parallel 
computation, the order in which the elements will be processed cannot be 
pre-determined. It is thus necessary for the correctness of the parallel execution 
that the update and assignment operations in the wave propagation phase are 
commutative and atomic. In practice, the order is not important for most
analysis methods based on the IWPP, including the example operations presented 
in this paper. 

\subsection{Parallel Queue Implementation and Management} \label{sec:parqueue}
A parallel queue is used to keep track of elements in the wavefront. 
An efficient parallel queue for GPUs is a challenging
problem~\cite{Hong:2011:ACG:1941553.1941590,DBLP:conf/memics/Karas10,Luo:2010:EGI:1837274.1837289}. 
A straight forward
implementation of a queue, as presented by Hong et
al.~\cite{Hong:2011:ACG:1941553.1941590}, could be done by employing an array
to store items in sequence and using atomic additions to calculate the position
where the item should be inserted, as presented in the code bellow.

\begin{verbatim}
  AddQueue(int* q_idx, type_t* q, type_t item) {
    int old_idx = AtomicAdd(q_idx, 1);
    q[old_idx] = item; }
\end{verbatim}

Hong et al. stated that this solution worked well for their use case. The use
of atomic operations, however, is very inefficient when the queue is heavily
employed as in our case. Moreover, a single queue that is shared among all
thread blocks introduces additional overheads to guarantee data consistency
across the entire device. It may also require inter-block
synchronization primitives~\cite{DBLP:conf/ipps/XiaoF10} to synchronize threads
from multiple block that are not standard methods supported by CUDA.

\begin{figure}[htb!]
\begin{center}
\includegraphics[width=0.6\textwidth]{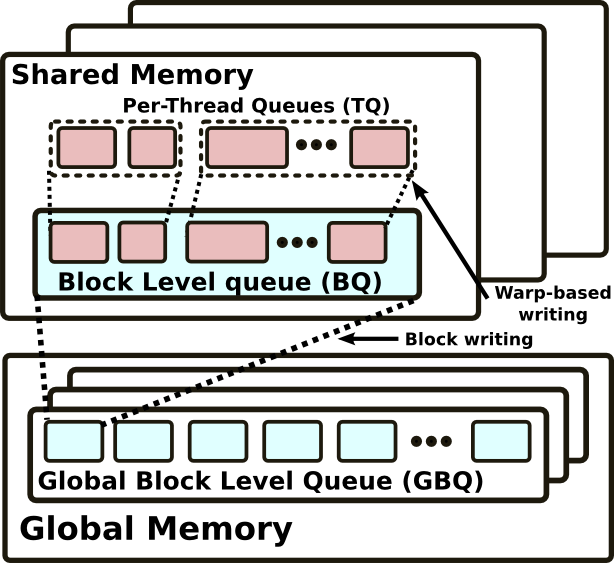}
\vspace*{-2ex}
\caption{Multi-level parallel queue.}
\vspace*{-2ex}
\label{fig:parallelqueue}
\end{center}
\end{figure}

To avoid these inefficiencies, we have designed a parallel queue that operates
independently in a per thread block basis to avoid inter-block communication. 
In order to exploit the fast memories in 
a GPU for fast write and read accesses, the queue is implemented as multiple 
levels of queues (as depicted in
Figure~\ref{fig:parallelqueue}):~(i)~Per-Thread queues (TQ) which are very small queues
private to each thread, residing in the shared memory; (ii)~Block Level Queue
(BQ) which is also in the shared memory, but is larger than TQ. Write operations 
to the block level queue are performed in thread warp-basis; and (iii)
Global Block Level Queue (GBQ) which is the largest queue and uses the global memory 
of the GPU to
accumulate data stored in BQ when the size of BQ exceeds the shared memory size.

In our queue implementation, each thread maintains an 
independent queue (TQ) that does not require any synchronization to be performed. 
In this way threads can efficiently store data elements from a given neighborhood in the 
queue for computation. Whenever this level of queue is full, it is necessary
to perform a warp level communication to aggregate items stored in the local
queues and write them to BQ. In this phase, a parallel thread warp
prefix-sum is performed to compute the total number of items queued in 
individual TQs. A single shared memory atomic operation is performed at the Block Level
Queue to identify the position where the warp of threads should write their
data. This position is returned, and the warp of threads write their local
queues to BQ.

Whenever a BQ is full, the threads in the corresponding thread block are
synchronized.  The current size of the queue is used in a single operation to
calculate the offset in GBQ from which the BQ should be stored. After this
step, all the threads in the thread block work collaboratively to copy in
parallel the contents of QB to GBQ. This data transfer operation is able to
achieve high throughput, because the memory accesses are coalesced. It should
be noted that, in all levels of the parallel queue, an array is used to hold
the queue content. While the contents of TQ and BQ are copied to the lower
level queues when full, GBQ does not have a lower level queue to which its
content can be copied. In the current implementation, the size of GBQ is
initialized with a fixed tunable memory space. If the limit of GBQ is reached
during an iteration of the algorithm, excess data elements are dropped and not stored.
The GPU method returns a boolean value to CPU to indicate that some data elements have
been dropped during the execution of the method kernel.  In that case, the
wavefront propagation algorithm has to be re-executed, but using the output of
the previous execution as input, because this output already holds a partial
solution of the problem. Recomputing using the output of the previous
step as input will result in the same solution as if the execution was carried
out in a single step. 

While adding new items to the queue requires a number of operations to
guarantee consistency, the read operations can be performed without introducing
synchronization. This is possible by partitioning input data elements
statically among threads, which keep consuming the data elements until all
items are computed. The drawbacks of a static partition of the queue and work
within a loop iteration are minimum, because the computation
costs of the queued data elements are similar; only a minor load imbalance is introduced
with this solution. Moreover, the queued data elements are repartitioned at each new
iteration of the loop. A high level description of the $dequeue$ function is
presented bellow.  In this function, the value of the item is initialized to a
default code corresponding to an empty queue. Then, the queue index
corresponding to the next item to be returned for that particular thread is
calculated, based on the size of the thread block ($block\_size$), the number
of iterations already computed by this thread warp ($iter$), and the identifier
of that thread in the block.

\begin{verbatim}
type_t DeQueue(int q_size, type_t* q, int iter) {
  type_t item = QUEUE_EMPTY;
  queue_idx = tid_in_block + iter * block_size;
  if(queue_idx < q_size){
     type_t item = q[queue_idx];
  }
  return item; }
\end{verbatim}

\subsection{GPU-enabled Fast Hybrid Morphological Reconstruction}

In this section, we present how the GPU-based
version of the Fast Hybrid Reconstruction algorithm, referred to in this paper
as FH\_GPU, is implemented (see Algorithm~\ref{alg:fhgpu}). 

The first stage of FH, as is described in Section~\ref{mr-algo}, consists of
the raster and anti-raster scans of the image. This stage has a regular computation 
pattern that operates on all pixels of the image. An efficient GPU implementation 
of the SR algorithm, including the raster and anti-raster scan stage, has been 
done by Pavel Karas~\cite{DBLP:conf/memics/Karas10}. This implementation is 
referred to as SR\_GPU in this paper. We employ the SR\_GPU implementation to 
perform the first phase of FH on a GPU. 
Note that the raster scan is decomposed into
scans along each dimension of the image.  The neighborhoods are defined as
$N_G^{+,0}$ and $N_G^{0,+}$ for the sets of neighbors for the row-wise and
column-wise scans, respectively.  Similarly, for the anti-raster scan,
$N_G^{-,0}$ and $N_G^{0,-}$ are defined for the respective axis-aligned
component scans.

\begin{algorithm}
\caption{Fast Parallel Hybrid Reconstruction}
\label{alg:fhgpu}
{\bf Input}
\vspace{-2.5mm}
\begin{description}
	\item[$I$:]      \emph{mask image}
	\vspace{-2.5mm}
	\item[$J$:]      \emph{marker image, defined on domain $D_{I}, J \le I$.}
\end{description}
\vspace{-2.5mm}
\begin{algorithmic}[1]
	\STATE \{\textbf{Initialization phase}\}
	\STATE Scan $D_I$ in raster order.
		\STATE{\ \ \bf for all\ } {$rows \in D_{I}$} {\bf do in parallel}
			\STATE{\ \ \ \ Let $p$ be the current pixel}
			\STATE{\ \ \ \ $J(p) \leftarrow (max\{J(q), q \in N_G^{+,0}(p) \cup \{p\}\}) \wedge I(p)$}
		\STATE{\ \ \bf for all\ } {$columns \in D_{I}$} {\bf do in parallel}
			\STATE{\ \ \ Let $p$ be the current pixel}
			\STATE{\ \ \ $J(p) \leftarrow (max\{J(q), q \in N_G^{0,+}(p) \cup \{p\}\}) \wedge I(p)$}

	\STATE Scan $D_I$ in anti-raster order.
		\STATE{\ \ \bf for all\ } {$rows \in D_{I}$} {\bf do in parallel}
			\STATE{\ \ \ \ Let $p$ be the current pixel}
			\STATE{\ \ \ \ $J(p) \leftarrow (max\{J(q), q \in N_G^{-,0}(p) \cup \{p\}\}) \wedge I(p)$}
		\STATE{\ \ \bf for all\ } {$columns \in D_{I}$} {\bf do in parallel}
			\STATE{\ \ \ Let $p$ be the current pixel}
			\STATE{\ \ \ $J(p) \leftarrow (max\{J(q), q \in N_G^{0,-}(p) \cup \{p\}\}) \wedge I(p)$}

	\STATE{\ \ \bf for all\ }{$p \in D_I$} {\bf do in parallel}
		\STATE {\ \ \ \ \bf if\ }{$\exists q\in N_G(p)\ |\ J(q)<J(p)\ and\ J(q)<I(q)$} 
		\STATE {\ \ \ \ \ \ queue\_add(p)}
\STATE \{\textbf{Wavefront propagation phase}\}
\WHILE{queue\_empty() = false}
	\FORALL{$p \in queue$} {\bf in parallel}
		\FORALL{$q \in N_{G}(p)$} {\bf in parallel}
			\IF{$J(q) < J(p)$ and $I(q) \neq J(q)$}
				\STATE $oldval = atomicMax(\&J(q), min\{J(p),I(q)\})$
				\IF{$oldval <  min\{J(p),I(q)\}$}
					\STATE queue\_add(q)
				\ENDIF
			\ENDIF
		\ENDFOR
	\ENDFOR
\ENDWHILE
\end{algorithmic}
\end{algorithm}

The transition from the raster scan phase to the wavefront propagation phase
is slightly different from the sequential FH. The SR\_GPU does not guarantee
consistency when a pixel value is updated in parallel. Therefore, after the
raster scan phase, pixels inserted into the queue for computation in the next
phase are not only those from the anti-raster neighborhood ($N_G^-$) as in the
sequential algorithm, but also pixels from the entire neighborhood ($N_G$)
satisfying the propagation condition (See Algorithm~\ref{alg:fhgpu}, Lines 16--18).

When two queued pixels (e.g., $p'\ and\ p''$) are computed in parallel, it is
possible that they have common neighbors ($N_G(p') \cap N_G(p'') \neq
\emptyset$). This can potentially create a race condition when updating the
same neighbor $q$. Updating the value of a pixel $q$ is done via a maximum
operation, which can be efficiently implemented to avoid race conditions by
employing atomic $Compare$-$and$-$Swap$ (CAS) based instructions ($atomicMax$
for GPUs). The value of pixel $q$ before the update ($oldval$) is returned,
which is used in our algorithm to test if $q$ was updated due to the
computation of the current pixel $p'$. If the $oldval$ is not smaller than
$min\{J(p'),I(q)\}$, it indicates that the processing of another pixel $p''$
updated the value of $q$ to a value greater than or equal to
$min\{J(p'),I(q)\}$. In this case, $q$ has already been queued due to the $p''$
computation and does not need to be queued again. Otherwise, $q$ is queued for
computation. The overall process continues until the queue is empty. Note that
the update operation is a maximum, hence an $atomicMax$ is used instead of the
$atomicCAS$ operation in the generic wavefront propagation skeleton. It leads
to better performance as the repeat-until loop in
Algorithm~\ref{alg:queuephasegpu} (Lines 5--14) is avoided. We also refer the
reader to a previous technical report~\cite{morph-recon} with additional
details on our GPU-based implementation of the Morphological Reconstruction.

\subsection{GPU-enabled Euclidean Distance Transform}
The GPU implementation of distance transform is shown in
Algorithm~\ref{alg:dtgpu}. The implementation is very similar to the sequential
version. The initialization phase assigns initial value to the Voronoi diagram
and adds contour pixels or those background pixels with foreground neighbors
for propagation (Lines~2~to~5). 

\begin{algorithm}
\caption{Distance Transform Algorithm for GPUs}
\label{alg:dtgpu}
\begin{description}
	\item[{\bf Input:}$I$:]      \emph{mask image}
\vspace{-2.5mm}
	\item[{\bf Output:}$M$:]      \emph{distance map image}
\vspace{-2.5mm}
\end{description}
\begin{description}
	\item[$FG$:] \emph{Foreground}; $BG$: \emph{Background}
\end{description}
\vspace{-2.5mm}
\begin{algorithmic}[1]
\STATE \{{\bf Initialization phase}\}
	\STATE{\bf for all\ }{$p \in D_I$} {\bf do in parallel}
		\STATE{\ \ $VR(p) = (I(p)==BG)\ ?\ p\ :\ inf$}
		\STATE {\ \ \bf if\ }{$I(p)==BG\ and\ \exists q\in N_G(p)\ |\ I(q)==FG$} 
		\STATE {\ \ \ \ queue\_add(p)}
\STATE \{{\bf Wavefront propagation phase}\}
\WHILE{queue\_empty() = false}
	\FORALL{$p \in queue$} {\bf in parallel}
		\FORALL{$q \in N_{G}(p)$} 
			\REPEAT
			\STATE $curVRQ = VR(q)$
			\IF{DIST$(q, VR(p) <$ DIST$(q, curVRQ)$}
				\STATE $old = atomicCAS(\&VR(q), curVRQ, VR(p))$
				\IF{$old \neq curVRQ$}
					\STATE queue\_add(q)
					\STATE $break$;
				\ENDIF
			\ELSE
				\STATE $break$
			\ENDIF
			\UNTIL{\texttt{True}}
		\ENDFOR
	\ENDFOR
\ENDWHILE
\STATE{\bf for all\ }{$p \in D_I$} {\bf do in parallel}
\STATE{\ \ $M(p) = $\ DIST$(p, VR(p))$}
\end{algorithmic}
\end{algorithm}

In the wavefront propagation phase, for each pixel $p$ queued for computation,
its neighbor pixels $q$ are checked to verify if using $VR(p)$ as path to a
background pixel would result in a shorter distance  as compared to the current
path stored in $VR(q)$. If it does, $VR(q)$ is updated and $q$ is queued for
computation. Once again, concurrent computation of pixels in the queue may
create a race condition if multiple threads try to update the same $VR(q)$.  To
prevent this condition, a compare and swap operation is employed when updating
$VR$. $VR$ is only updated if the value used to calculated distance ($curVRQ$)
is still the value stored in $VR(q)$.  Otherwise, the algorithm reads the new
value stored in $VR(q)$ and calculates the distance again (Line~12).  The final
step, as in the sequential case, computes the distance map from $VR$.  It is
performed in parallel for all points in the input image.

A given VR diagram may have multiple valid solutions. If two pixels, $p'$ and
$p''$, are equidistant and the closest background pixels to a foreground pixel
$q$, either of the pixels can be used as the closest background
pixel of $q$ ($VR(q) = p'$ or $VR(q) = p''$). In other words, a 
foreground pixel could initially be assigned one of multiple background pixels, as long
as the distance from the foreground pixel to those background pixels is the
same. In our algorithm, the order in which the wave propagations are computed
defines whether $p'$ or $p''$ are used in such a case.  However, since both
pixels are equidistant to $q$, using either of them would result in the same
distance map. Thus, the results achieved by the parallel distance transform are
the same as those of the sequential version of the algorithm.

\section{Parallel Execution on Multiple GPUs and CPUs}
\label{sec:multigpu}
\begin{figure*}[htb!]
\begin{center}
        \includegraphics[width=0.95\textwidth]{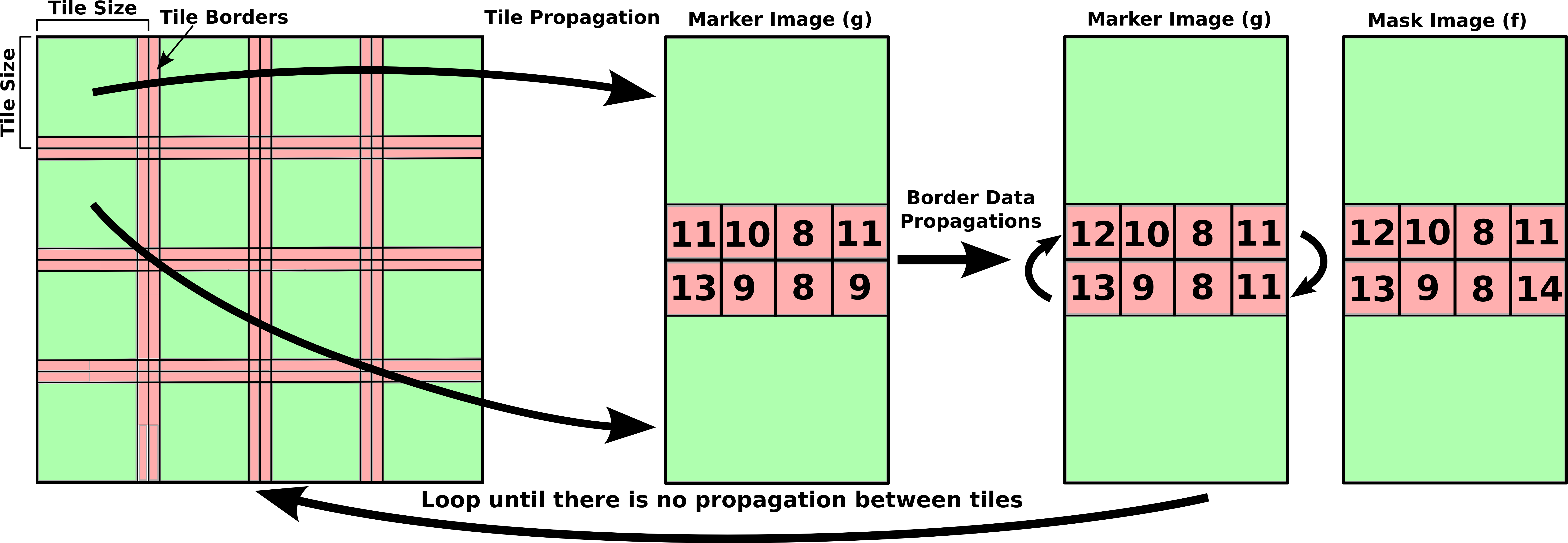}
\vspace*{-2ex}
\caption{Tiling based parallelization (tiles are not necessarily square).}
\vspace*{-3ex}
\label{fig:parallelStrategy}
\end{center}
\end{figure*}

Our parallelization strategy, shown in
Figure~\ref{fig:parallelStrategy}, for multiple GPUs and CPUs 
divides the input data domain into non-overlapping partitions (tiles). 
Local wavefront propagations are computed 
independently for each partition. After the local propagations are computed, 
border information is exchanged to allow for inter-partition propagations. 
This process continues until there are no more propagations within a partition 
and between partitions.

Inter-processor communication is necessary because the irregular wavefront 
propagation pattern involves wavefront propagations that are data dependent. 
The shape of the wavefront and the amount of wave propagation
cannot be estimated prior to execution. As a result, a partitioning 
of the input data domain across processing units can incur inter-partition 
propagations and inter-processor communication, which cannot be  
be eliminated by padding each domain partition.

\begin{figure*}[htb!]
\begin{center}
\mbox{
 \subfigure[Marker {\em g} and mask {\em f} images.]{\label{fig:initial} 
        \includegraphics[width=0.46\textwidth]{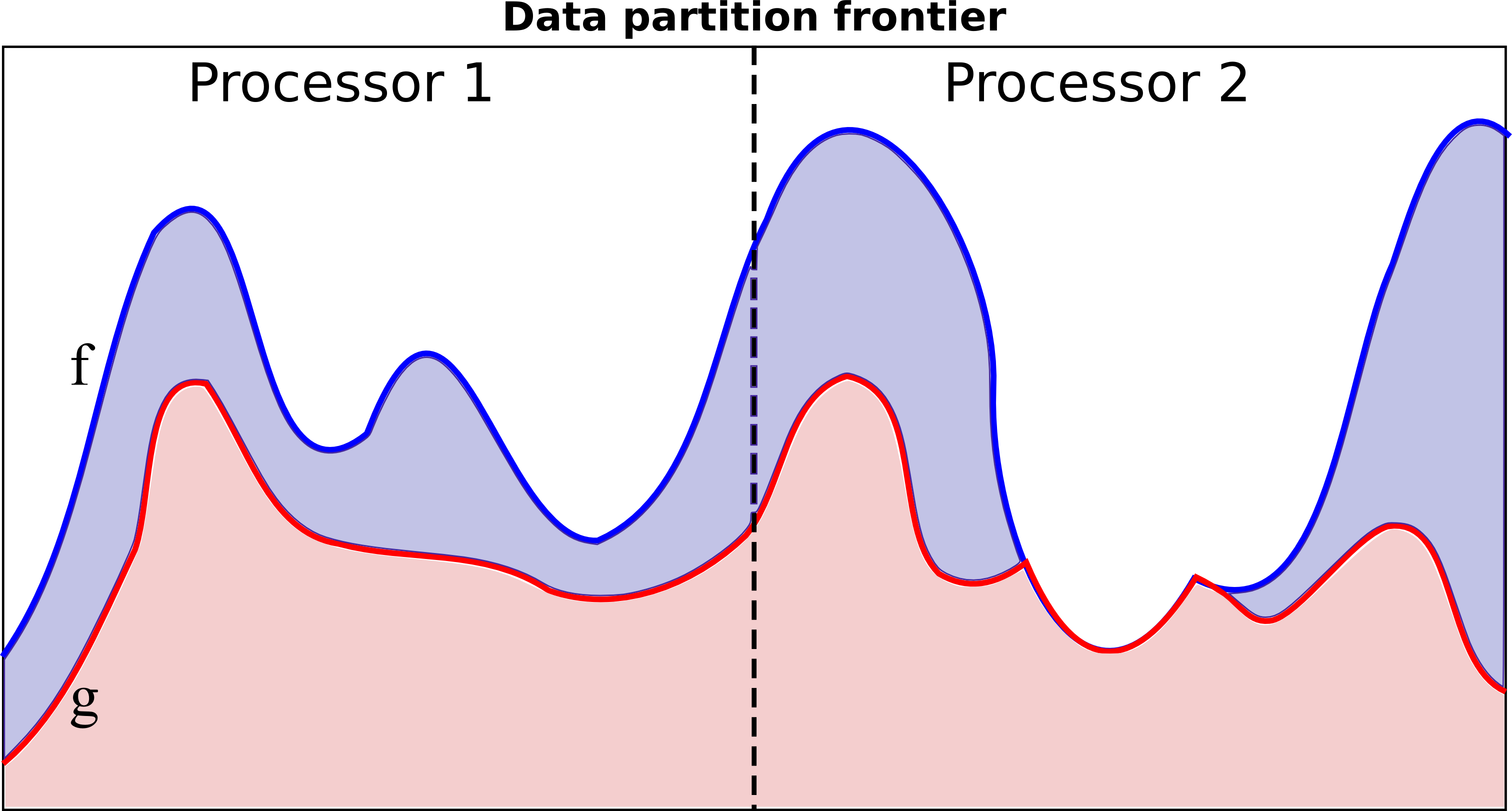}}
}
\mbox{
 \subfigure[Processor local reconstruction.]{\label{fig:local} 
        \includegraphics[width=0.46\textwidth]{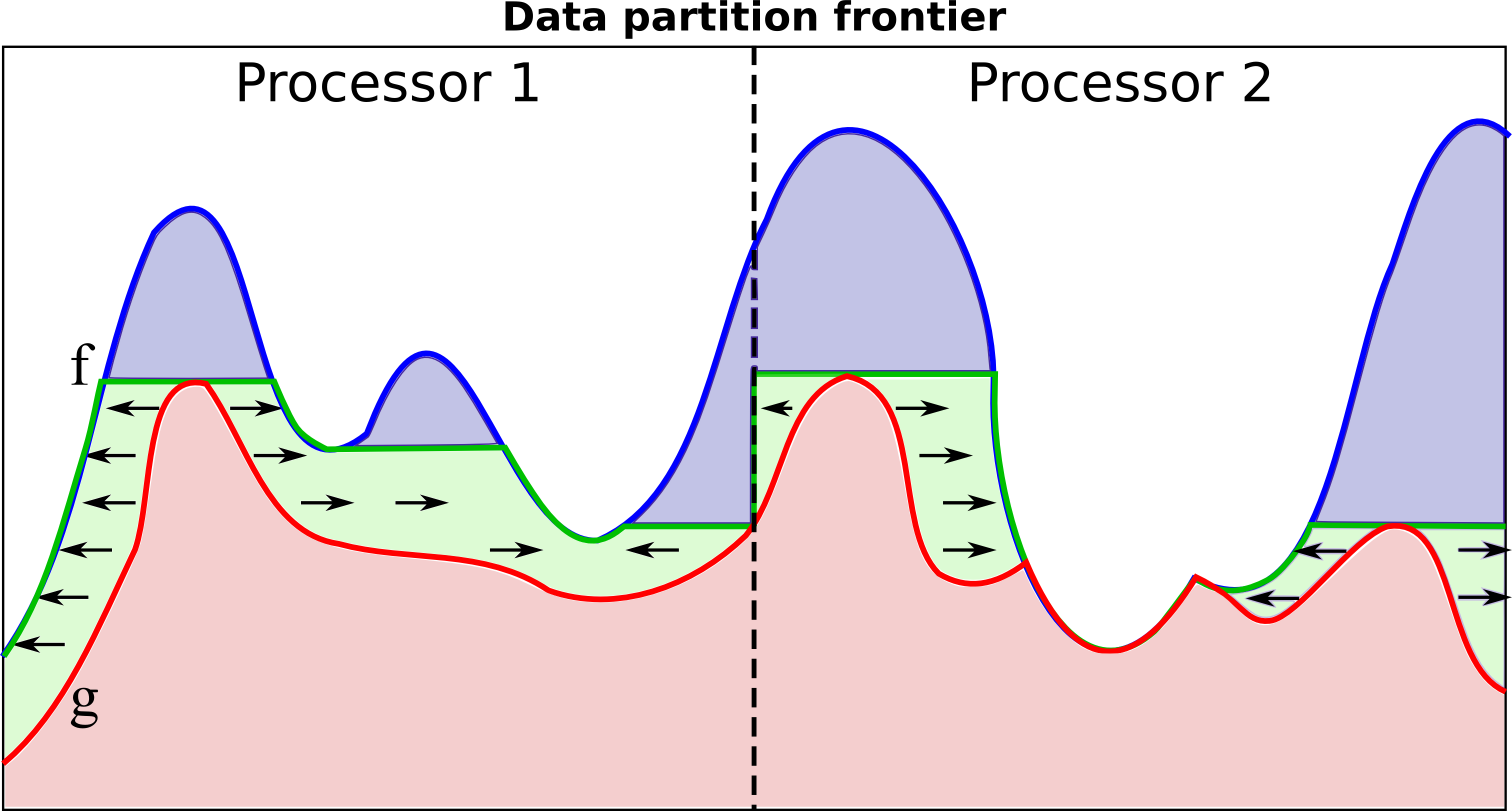}}
}
\mbox{
 \subfigure[Reconstruction after inter-processor propagation.]{\label{fig:global} 
        \includegraphics[width=0.6\textwidth]{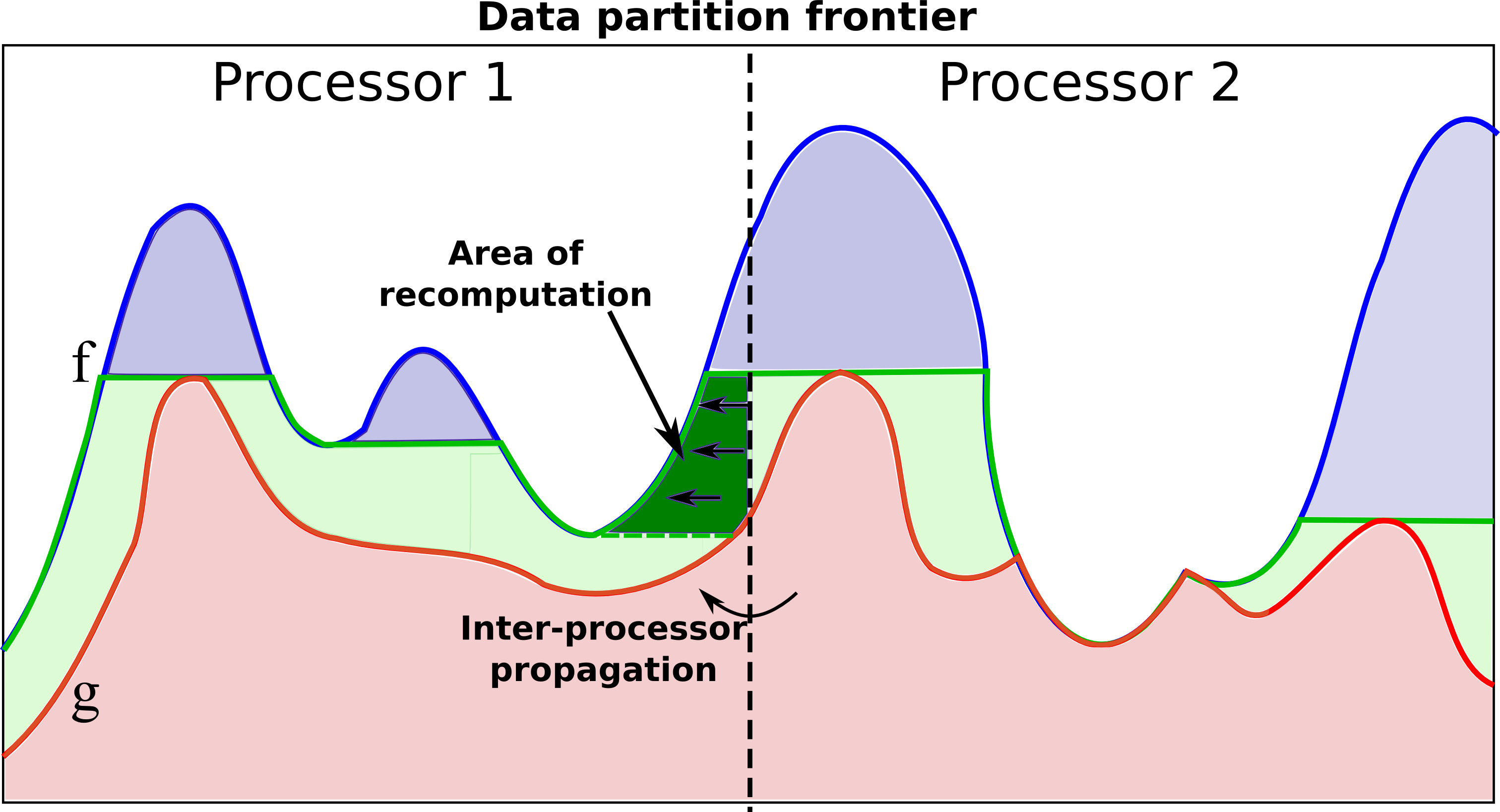}}
}

\vspace*{-2ex}
\caption{Evaluating Scalability According to the Number of Blocks.}
\vspace*{-3ex}
\label{fig:reconPartitionExample}
\end{center}
\end{figure*}

Figure~\ref{fig:reconPartitionExample} shows an example in which an input data
domain is partitioned between two processors. The morphological reconstruction
algorithm is used in this example.  The input data domain, described in a
1-dimensional space in the example, is equally divided between processors~1
and~2 without any replication at the beginning of the execution
(Figure~\ref{fig:initial}).  Each processor independently performs the
wavefront propagation in their respective partitions by identifying seeds and
computing the usual propagation as required by the algorithm.  When the
stability is reached (i.e., the queue is empty) in the partitions, there 
may be propagations crossing partition boundaries that have not been computed.
Figure~\ref{fig:local} presents the result of propagations within a partition. 
To compute inter-partition propagation effects correctly, processors
must exchange border information to identify inter-partition propagations. After 
border information is exchanged, propagations originating from partition borders 
are initialized, and another round
of local propagations is executed. The process of performing local propagations
and exchanging border information continues until no more propagations are
possible within and between partitions.  The final result of the
morphological reconstruction for the 1-dimensional example is presented in
Figure~\ref{fig:global}. The area recomputed because of the inter-partition
propagations is represented in dark green.

\begin{figure*}[htb!]
\begin{center}
        \includegraphics[width=0.7\textwidth]{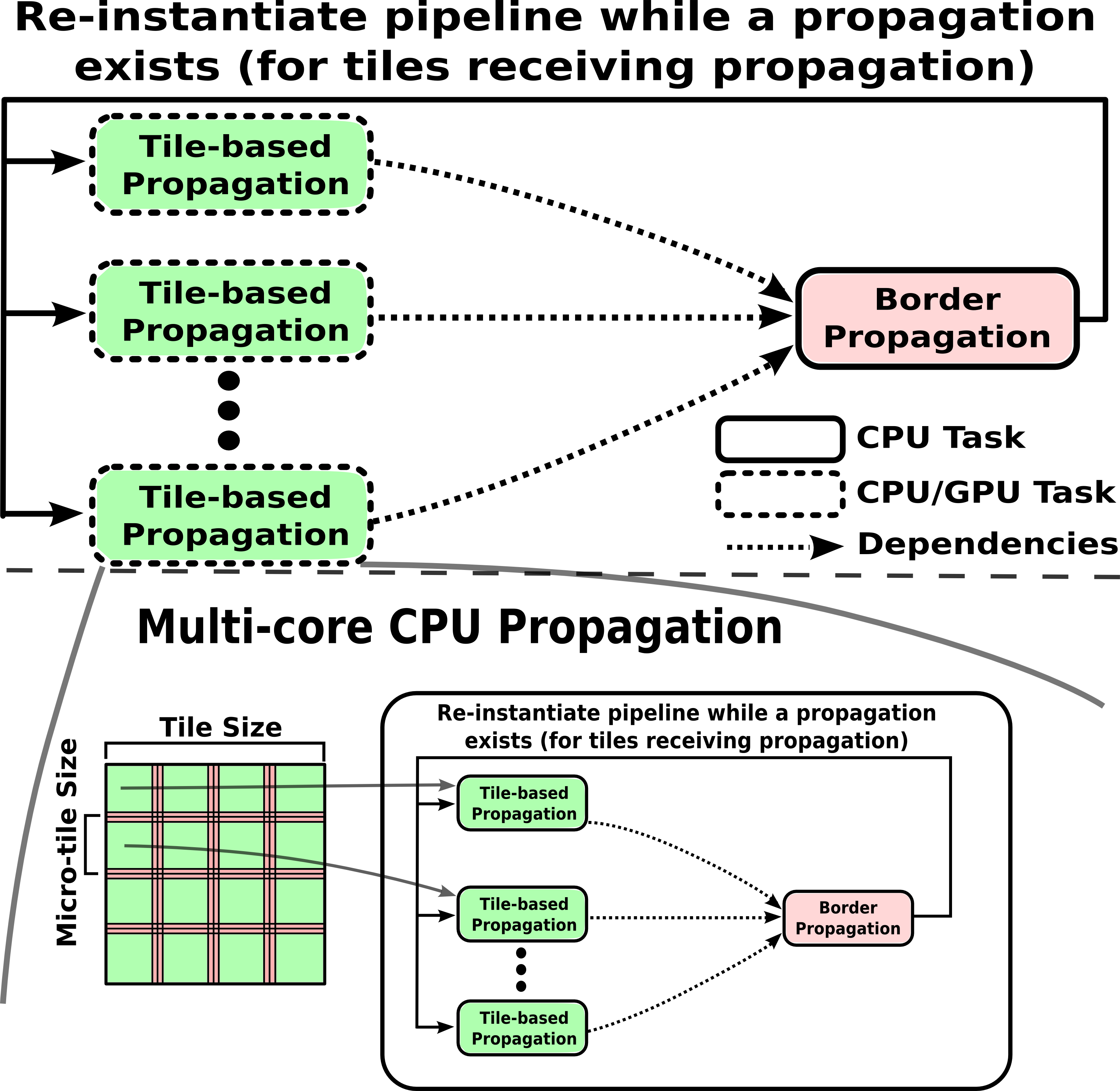}
\vspace*{-2ex}
\caption{Pipeline representation of the CPU-GPU wavefront propagation.
Instances of Tile-based Propagation (TP) compute
propagation in each tile as shown at the top. 
A Border Propagation (BP) stage, which depends
on TP instances for execution, is also created to resolve propagation among the tiles. If a
propagation exists between tiles, BP re-instantiates the pipeline. This process 
continues until stability is reached. The multi-level tiling strategy, in which 
a tile is repartitioned into micro-tiles for multi-core execution, is shown at 
the bottom. This strategy is implemented to reduce CPU vs GPU computational load imbalance.}
\vspace*{-3ex}
\label{fig:pipelineParallel}
\end{center}
\end{figure*}

The parallel implementation consists of a pipeline of two stages
(Figure~\ref{fig:pipelineParallel}): Tile Propagation (TP) and Border
Propagation (BP). The TP stage performs local propagations in tiles. 
An instance of this stage is created for each tile. 
The BP stage, on the other hand, computes propagations from data elements 
at tile borders. In essence, BP transmits
propagations from one tile to another. When there is a need for
propagation between tiles, BP will re-instantiate the pipeline.  In that case,
a TP instance is created for each tile receiving propagation and scheduled for
execution. It computes 
propagations within the tile that are initiated from the borders.
A new instance of BP is also created with a dependency on the new TP instances. That is, 
it is not scheduled for execution until all of the new TP instances finish their propagation 
computations. This cycle is repeated 
until there are no more intra-tile and inter-tile propagations. 

The instances of the two stages are
dispatched for execution on available CPUs and GPUs, as long as a stage
has an implementation for those devices. We employ the concept of function
variant, which is a group of functions with same name, arguments, and result
types~\cite{merge,Millstein:2004:PPD:1035292.1029006}. In our implementation a
function variant for a task is the CPU and GPU implementations of the
operation. Binding to a function variant enables the runtime system to choose
the appropriate function or functions during execution, allowing multiple
computing devices to be used concurrently and in a coordinated manner. We should 
note that a function variant provides opportunities for runtime optimizations, but 
an application is not required to provide CPU and GPU implementations for all of 
the stages.

\begin{figure*}[htb!]
\begin{center}
        \includegraphics[width=0.6\textwidth]{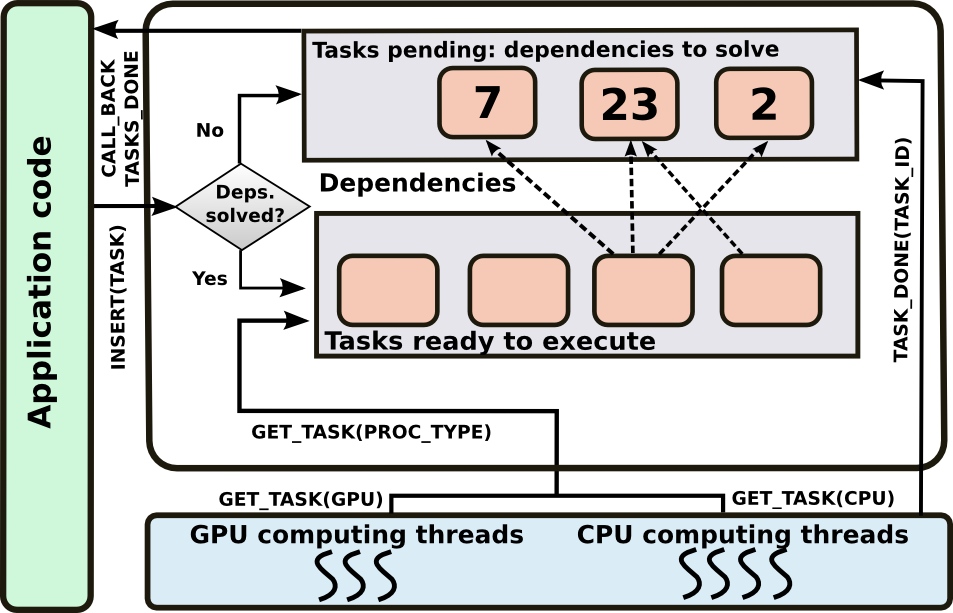}
\vspace*{-2ex}
\caption{System support for execution of pipeline of tasks in machines equipped with multi-core CPUs and GPUs.}
\vspace*{-3ex}
\label{fig:resourceManager}
\end{center}
\end{figure*}

Figure~\ref{fig:resourceManager} presents an overview of our system
support~\cite{pipeline-hierarchical,Teodoro-IPDPS2012} to schedule pipeline
applications in machines equipped with multiple CPUs and GPUs.  Tasks (TP and BP 
instances) dispatched for execution by the wavefront propagation pipeline
are either queued as ready to execute or inserted in a queue of pending tasks,
respectively, depending on whether their dependencies are resolved or not.
Tasks ready for execution are consumed by computing threads responsible for
managing CPU cores and GPUs.  A computing thread is assigned to each CPU-core
or GPU available in the default mode. Assignment of tasks to computing devices
is done in a demand-driven manner.  When a CPU core or GPU remains idle, one of
the tasks is assigned to the idle device.  The default scheduling policy for
choosing the task to be execute into a device requesting work is FCFS (first
come, first served). In the case of our application, since the BP instance will
be dependent on all TP instances, the BP is only dispatched for execution
after all the TPs have finished the execution.

When a TP instance is scheduled for GPU execution, the tile to be processed is
copied to the GPU memory. The results are transferred back to the CPU memory
at the end of the processing. TP instances are dynamically assigned to processing 
devices using a demand-driven approach to alleviate load imbalance
that may exist due to different computation requirements among tiles.
Moreover, with a partitioning of input data into tiles, GPUs can be used when
the input data does not fit into GPU memory as long as individual 
tiles do.

Our experimental evaluation showed that the maximum GPU performance for the
IWPP is achieved when large input domain partitions are used. This is also
observed for other types of GPU computations as larger inputs usually lead to
higher computation times that better offset overheads of using GPUs. However,
when CPUs and GPUs are used together, large tiles tend to reduce the
performance gain because of the greater gap in execution times on a CPU core vs
on a GPU. This increases load imbalance between the CPU and the GPU. To
reduce the impact of load imbalance, our implementation allows for parallel
CPU processing variants to be used. In this approach, instead of scheduling a
single partition per CPU core, a single partition is scheduled to a group of
CPU cores, if a parallel variant of the CPU processing operation is available.
Consequently, a partition assigned to a CPU will have an execution
time closer to that of a partition assigned to a GPU. 

In the wavefront propagation pattern, as presented on the bottom in
Figure~\ref{fig:pipelineParallel}, a partition assigned for parallel CPU
computation is broken into smaller micro-partitions that are processed by
multiple CPU cores in parallel. Border communication between partitions
assigned to different CPU cores is performed for inter-partition propagations.
The runtime system (Figure~\ref{fig:resourceManager}) is responsible for
setting the number of CPU cores available for a given partition. The multi-core
version of the propagation stages is implemented on top of
OpenMP~\cite{openmp}.

\section{Experimental Results} \label{sec:results}
\begin{figure}[htb!]
\begin{center}
\includegraphics[width=0.7\textwidth]{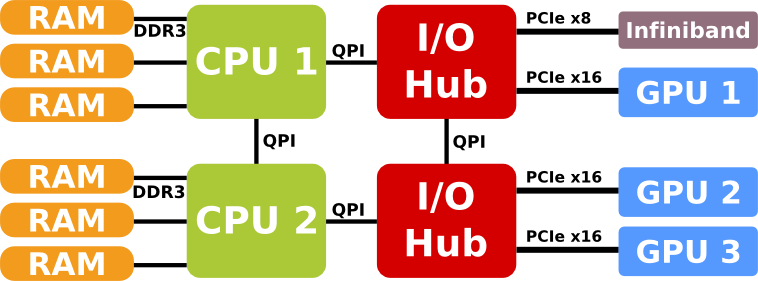}
\caption{Hybrid, Multi-GPU computing node.}
\label{fig:node}
\end{center}
\end{figure}

We evaluate the performance of the irregular wavefront 
propagation implementation using the morphological reconstruction and distance transform 
operations. These operations 
were applied on high-resolution RGB images from datasets collected 
by research studies in the In Silico Brain Tumor Research Center (ISBTRC)~\cite{insilico} at 
Emory University. The experiments were carried out on a machine (a single computation node of the 
Keeneland cluster~\cite{10.1109/MCSE.2011.83}) with 
Intel\textsuperscript{\textregistered} Xeon\textsuperscript{\textregistered} 
X5660 2.8 GHz CPUs, 3 NVIDIA Tesla M2090 (Fermi) GPUs, and 24GB of DDR3 RAM 
(See Figure~\ref{fig:node}). A total of 12
CPU computing cores available because hyper-threading mechanism is not enabled.
Codes used in our evaluation were compiled with ``gcc 4.2.1'', ``-O3''
optimization flag, and NVidia CUDA SDK 4.0. 
The experiments were repeated 3
times, and, unless stated, the standard deviation was not observed to be higher
than 1\%.  The speedup values presented in the following sections are 
calculated based on the performance of single CPU-core versions of 
the operations.
\subsection{Multi-core CPU Parallelization}
Two different multi-core CPU parallelization strategies were implemented 
on top of OpenMP: (i)~\emph{Tile-based parallelization} that partitions 
an input image into
regions and iteratively solves propagations local to each partition and
performs inter-region propagations, as described in Section~\ref{sec:multigpu};
(ii)~\emph{Non-Tiled parallelization} is an alternative parallelization in which 
points in the initial
wavefront are distributed in round-robin among CPU threads at the beginning
of the execution. Each thread then independently computes propagations
associated with the points assigned to it. To avoid inter-thread interference
in propagations, it is necessary to use $atomic$ compare-and-swap operations
during data updates.

\begin{figure}[htb!]
\begin{center}
\includegraphics[width=0.7\textwidth]{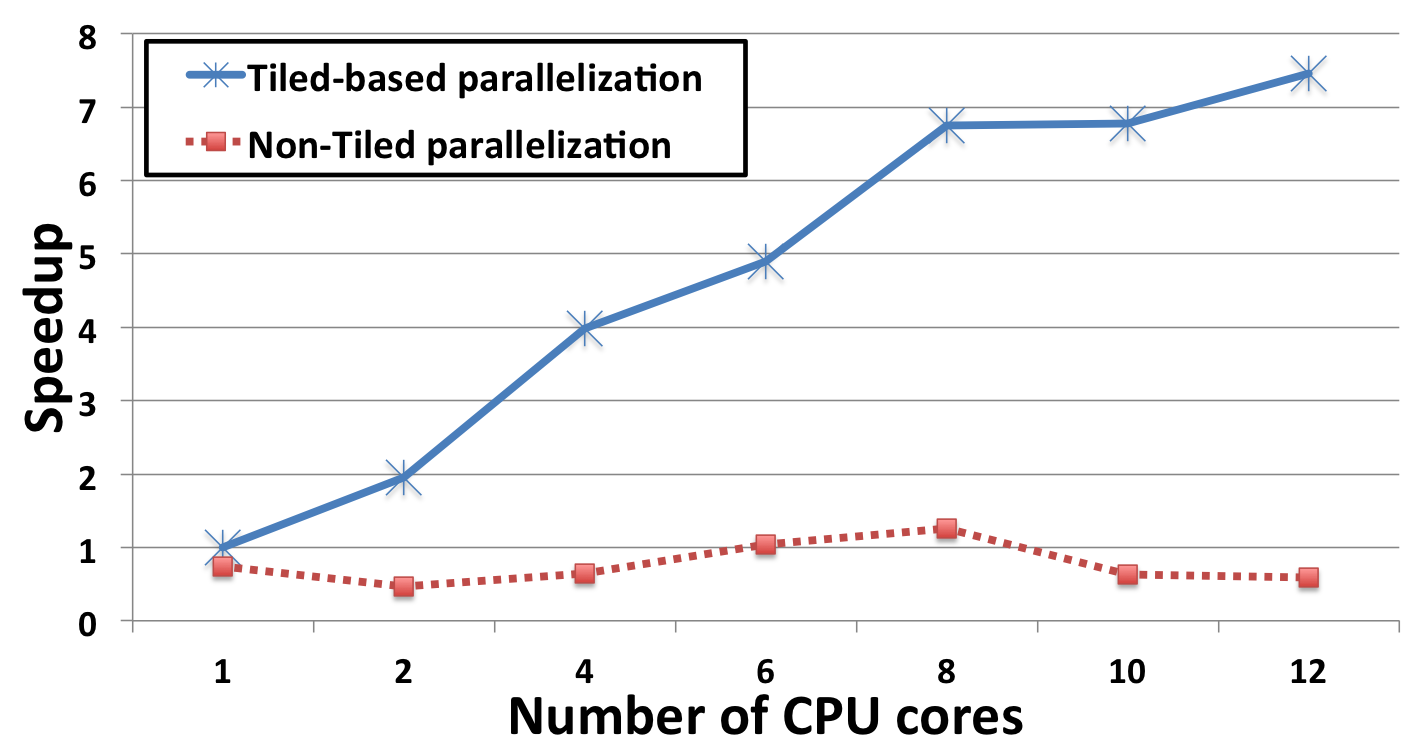}
\vspace{-2ex}
\caption{Multicore performance of morphological reconstruction. Different
parallelization strategies are compared.}
\label{fig:multicore-parallelization}
\end{center}
\vspace{-2ex}
\end{figure}

The performances of both parallelization strategies are presented in
Figure~\ref{fig:multicore-parallelization}. The non-tiled parallelization of
the morphological reconstruction resulted in slowdown as compared to the
sequential algorithm for most of the configurations; only a slight
improvement was observed on 8 CPU cores. In order to understand this
poor performance, the algorithm was profiled using \emph{perf} profiling
tool~\cite{perf} to measure the effects of increasing number of threads to
cache and bus cycles. Additional bus cycles are used for data read and write 
as well as for data transfers to
assert cache coherence among threads. 
An increase
in the number of threads from 1 to 2 in the non-tiled parallelization results 
in twice more cache misses and three
times more bus cycles. Similar behavior is observed as the number of threads
keep increasing, preventing the algorithm from having gains as more CPU cores
are used. For this parallelization, threads may be concurrently computing in
the same region of the input domain, and updates from one thread will
invalidate cache lines of all the other threads processing the same region. This
makes the cost of maintaining cache coherence very high.

The performance of the tile-based parallelization is much better with a 
speedup of $7.5\times$ in the configuration with 12 threads. This is a result of avoiding inter-thread 
interference. Because threads are assigned
to independent partitions, they do not
modify the same regions and affect each other. Nevertheless, 
the speedup performance of the algorithm is sub-linear. This is an
inevitable consequence of the threads' having to share the processor 
cache space and the bus subsystem. Since the algorithm is very
data intensive, larger cache and fast access to memory are primordial to
achieve maximum performance, and contention on these resources limits 
gains in CPU multicore systems. 
\subsection{GPU Parallelization}
The experiments in this section examine the impact on performance of the queue 
design as well as tile size and input data characteristics for the GPU-enabled 
implementations of irregular wavefront propagation computations.  
\subsubsection{Impact of Queue Design Choices}
This set of experiments evaluates the performance of three different parallel 
queue approaches: (i)~Na\"{\i}ve. This is the simple queue implementation based on
the atomic add operations only; (ii)~Prefix-sum (PF). PF performs a prefix-sum
among GPU threads to calculate the positions where items are stored in the
queue, reducing the synchronization cost due to the atomic operations; and,
(iii)~Prefix-sum + Per-Thread Queue (TQ). TQ employs local queues in addition
to the prefix-sum optimization in order to reduce the number of
prefix-sum/synchronization phases. In the experiments we present results from 
the morphological reconstruction implementation -- there is no significant 
difference between these results and those from the distance transform 
implementation.

The different queue versions are compared using a 4-connected grid and a
4K$\times$4K input image tile with 100\% of tissue coverage.  As a strategy to
initialize the queue with different number of pixels and evaluate its
performance under various scenarios, the number of raster and anti-raster scans
performed in the initialization is varied.  A single thread block is created
per Stream Multiprocessor (SM) of the GPU.  The execution times reported in
Table~\ref{tab:queue} correspond to the wavefront phase only. 

\begin{table}[htb!]
\begin{center}
\begin{tabular}{c c c c c c}
\hline
Initial & Total & \#Raster Scans & Na\"{\i}ve 	& PF 	& 	TQ 	 \\ \hline \hline
534K & 30.4M & 7	&	685	&  291	&	228	 \\ \hline 
388K & 22.2M & 9	&	527	&  223	&	175	 \\ \hline 
282K & 16.7M & 11	&	409	&  164	&	134	 \\ \hline 
210K & 12.7M & 13	&	323	&  133	&	106	 \\ \hline 
158K & 9.9M & 15	&	255	&  118	&	98	 \\ \hline 
124K & 7.7M & 17	&	207	&  99	&	83	 \\ \hline 
97K & 6.2M & 19	&	162	&  84	&	70	 \\ \hline
\end{tabular}
\end{center}
\vspace{-6mm}
\caption{Execution times in milliseconds (ms) for different versions of the queue: 
Na\"{\i}ve, Prefix-sum (PF), and Prefix-sum + Per-Thread Queue (TQ). The columns Initial 
and Total indicate the number of pixels initially queued before the wavefront propagation 
phase is executed and the total number of pixels queued/dequeued during the execution, 
respectively.}
\label{tab:queue}
\vspace*{-1ex}
\end{table}

The results show that the Na\"{\i}ve queue has the highest execution times in
all the experiments. The PF implementation was on average able to improve the
performance compared to Na\"{\i}ve version by 2.31$\times$. These performance
improvements indicate that atomic operations, while they are much faster in
modern GPUs~\cite{citeulike:8927897}, still introduce significant overhead and
that the heavy use of atomic operations should be considered carefully.  The
results in the table also show that TQ, which uses local queues to further
reduce synchronization, achieved performance improvements of 1.24$\times$ on
top of PF. These gains with the different versions of the queue
demonstrate the benefits of employing prefix-sum to reduce the number of atomic
operations and local queues to minimize synchronization among threads.
\subsubsection{Effects of Tile Size} \label{sec:input-size}
\begin{figure}[htb!]
\begin{center}
        \includegraphics[width=0.8\textwidth]{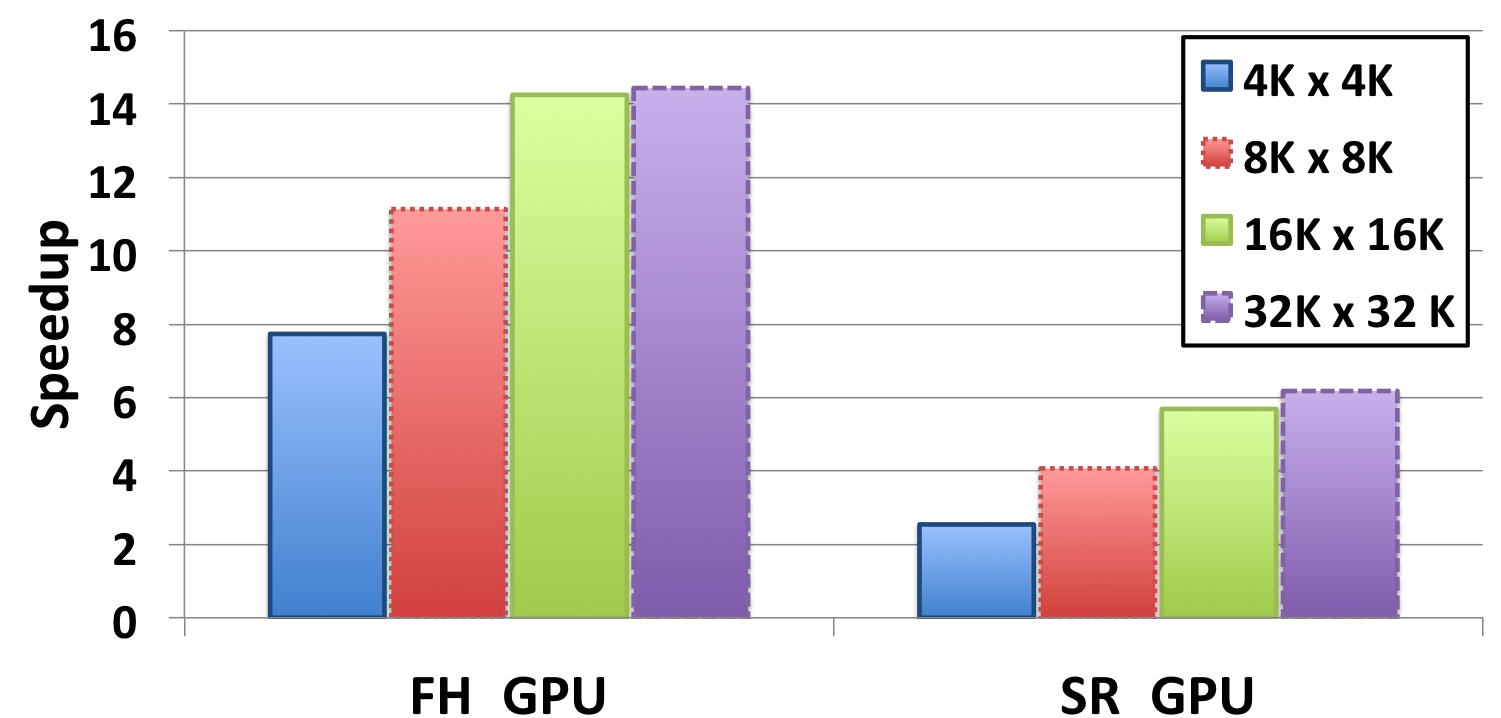}
\vspace{-2ex}
\caption{Morphological Reconstruction: performance of the IWPP based implementation (FH\_GPU) and Pavas algorithm~\cite{DBLP:conf/memics/Karas10} (SR\_GPU) that is implemented as iterative raster scanning according to the input data tile size.}
\label{fig:input-tilesize-morph}
\end{center}
\vspace{-2ex}
\end{figure}

The experiments in this section evaluate the impact of the tile size to the
performance of the morphological reconstruction and distance transform
algorithms. Figure~\ref{fig:input-tilesize-morph} presents the morphological
reconstruction algorithm performance using an 8-connected $G$ grid and an input image
of 64K$\times$64K pixels with 100\% tissue coverage. Two versions of this
algorithm were evaluated: FH\_GPU that is implemented using the irregular
wavefront propagation framework; and, SR\_GPU~\cite{DBLP:conf/memics/Karas10}
that is based on raster and anti-raster scans. SR\_GPU was the fastest GPU
implementation of morphological reconstruction available before our work. All
speedup values are calculated using the single core CPU version of the application 
as reference. 

The results show that the performances of both GPU algorithms improve as 
the tile size is increased until 
16K$\times$16K pixels; beyond this size no significant additional gains are 
observed. The better speedup values with larger tiles are the result of  
(i)~better amortization of initialization of computations with the GPU, 
specially in the raster scan pass (SR\_GPU and the initialization phase 
of FH\_GPU), which launches four GPU $kernels$ per iteration; and 
(ii)~the larger amount of parallelism available with big tiles. The experimental 
results show that FH\_GPU performs better than
SR\_GPU~\cite{DBLP:conf/memics/Karas10} in all the configurations, on average 
about 2.6$\times$ better. 
Further, FH\_GPU achieves a performance improvement of 14.3$\times$ compared to the single
core CPU version. It is important to highlight, however, that the FH\_GPU is a
two phase algorithm that includes raster scans in the initialization phase. The 
initial scan is responsible for about 43\% of the execution time, 
reducing performance gains from the wavefront propagation phase.
\begin{table}[htb!]
\begin{center}
\begin{tabular}{c c c c c}
\hline
Tile Size	& 	4K$\times$4K 	& 8K$\times$8K  & 16K$\times$16K & 32K$\times$32K \\ \hline \hline
Speedup		&	31.4		&  	36.8	&	37.5	 &	37.1	 \\ \hline 
\end{tabular}
\end{center}
\vspace{-4mm}
\caption{Distance Transform: performance according to the input data tile size.}
\label{tab:input-tilesize-distance}
\end{table}

The performance of the distance transform implementation is presented in 
Table~\ref{tab:input-tilesize-distance}. The initialization phase in this 
algorithm is inexpensive. Most of the execution time is spent in the wavefront 
propagation phase, which is less sensitive to the tile size because of its 
lower overheads -- all of the computation in this phase is performed within 
a single GPU $kernel$, while in the raster scanning phase 4 $kernel$ calls 
are executed per pass on the image.
\subsubsection{Impact of Input Data Characteristics} \label{sec:input-characteristics}
\begin{figure}[htb!]
\begin{center}
\includegraphics[width=0.7\textwidth]{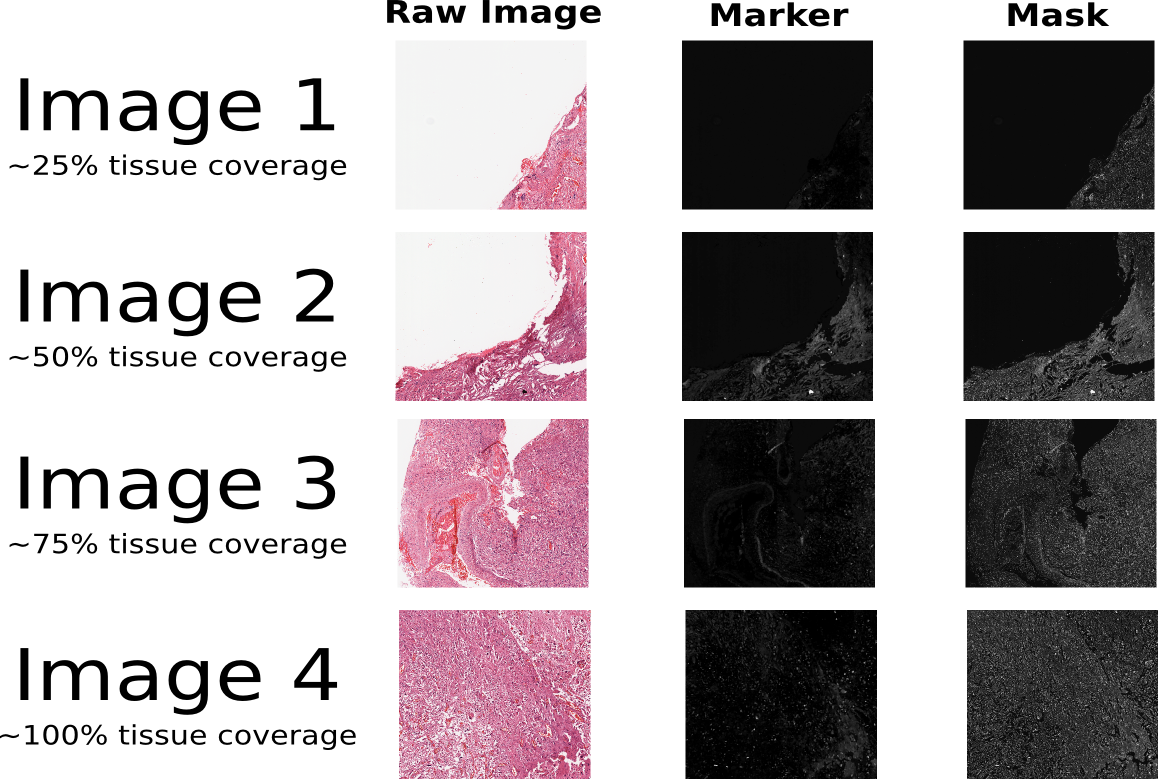}
\vspace{-2ex}
\caption{Raw image, marker, and mask for tiles with different amount of tissue coverage.}
\label{fig:tiles-coverage}
\end{center}
\vspace{-2ex}
\end{figure}

These experiments employed four different images with 64K$\times$64K pixels,
shown in Figure~\ref{fig:tiles-coverage}, that have different percentage of
tissue coverage. The 25\%, 50\%, 75\%, and 100\% of the image area is covered
with tissue (in contrast to background) in images 1, 2, 3, and 4, respectively.
The images are partitioned into tiles of size 16K$\times$16K to fit into the GPU
memory. Inter-partition propagations are performed as discussed in
Section~\ref{sec:multigpu}.

\begin{figure}[htb!]
\begin{center}
        \includegraphics[width=0.7\textwidth]{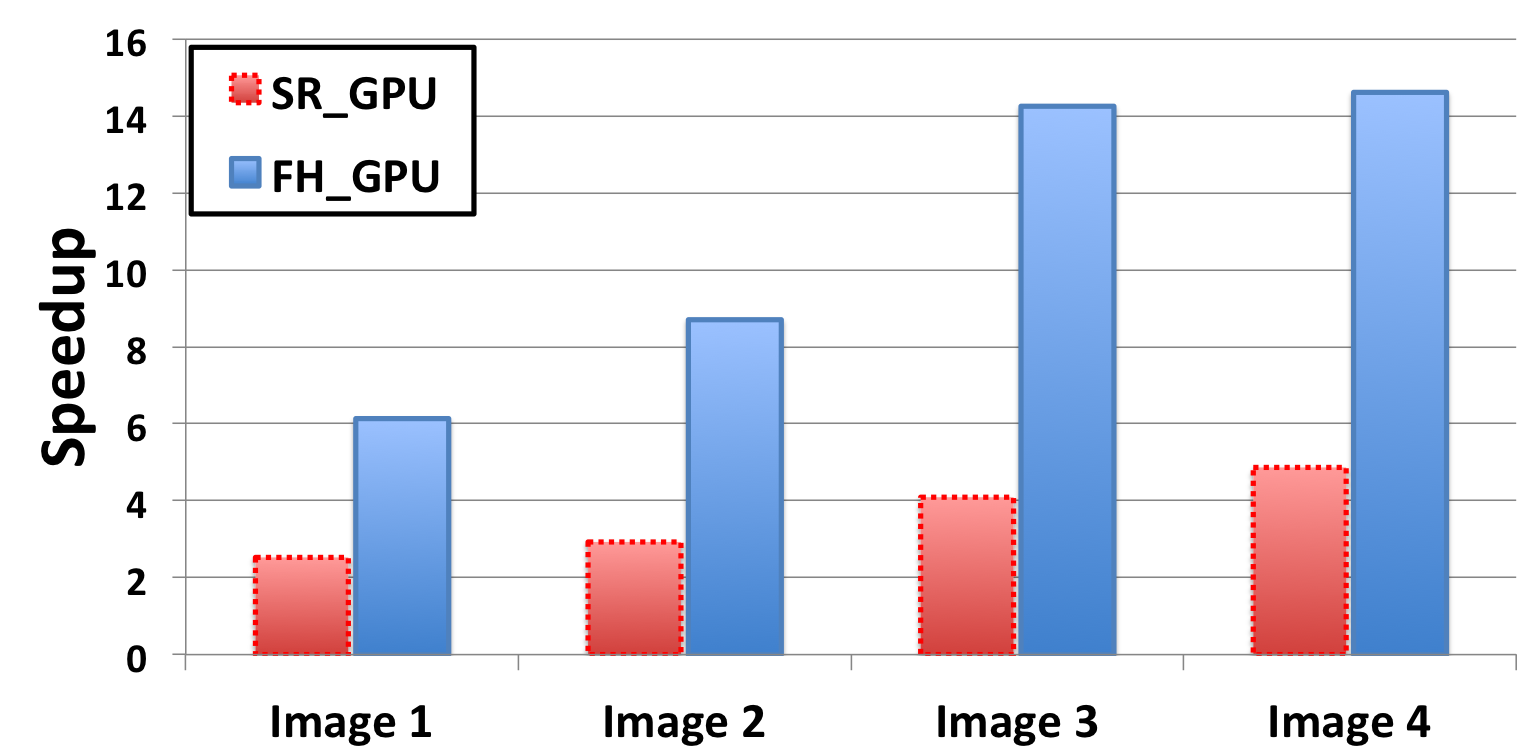}
\caption{Performance variation of the morphological reconstruction implementation 
with respect to input data characteristics.} 
\label{fig:input-characteristics}
\end{center}
\end{figure}

The speedup values of FH\_GPU and SR\_GPU with respect to the single core CPU
version are presented in Figure~\ref{fig:input-characteristics}. The
performances of the GPU implementations get better with images with larger tissue
coverage. Larger tissue coverage implies more work that can be done in parallel, which 
amortizes the overhead of starting the GPU computations. In addition, the higher 
parallelism for images with larger
tissue coverage increases the utilization of GPU threads and reduces the chances
of having conflicting atomic operations between threads. 

The comparison of FH\_GPU and SR\_GPU shows a consistent performance gap
between the implementations for all variations of the input images. 
In our experiments, FH\_GPU
achieved performance improvements of up to 3.1$\times$ on top of SR\_GPU. The
performance gains of FH\_GPU are higher for images with less
tissue coverage. For those images, SR\_GPU passes over the entire input image domain 
irrespective of the number of pixels being modified, resulting in more unnecessary 
computation that does not contribute to the output.

\begin{figure}[htb!]
\begin{center}
        \includegraphics[width=0.7\textwidth]{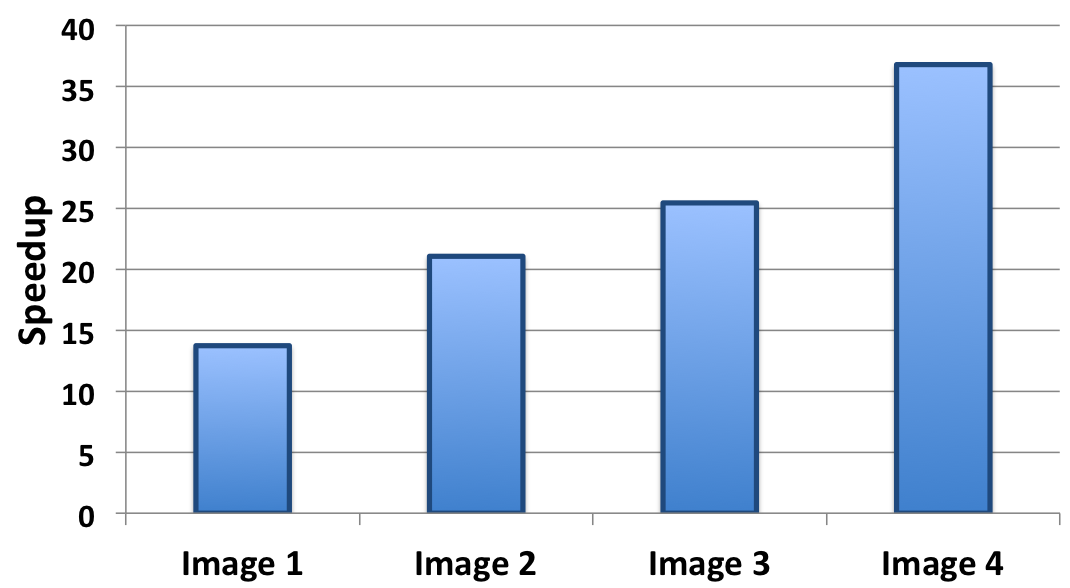}
\caption{Performance variation of the distance transform implementation with respect to input data
characteristics.}
\label{fig:input-characteristics-distance}
\end{center}
\end{figure}

The performance of the distance transform implementation as the input data is varied
is presented in Figure~\ref{fig:input-characteristics-distance}. As in the morphological 
reconstruction implementation, larger tissue coverage results in higher
speedup values. Once again, this is a consequence of more computation and
parallelism available with images with more tissue coverage. More tissue coverage means 
there are more foreground pixels from which to calculate distances 
to a background pixel.  
\subsubsection{Impact of Exceeding Queue Storage Limit}
This set of experiments vary the memory space available for the queue to store
pixels. When the queue storage limit is exceeded, our implementation throws away
excess pixels. Then it is necessary to perform another round of computations to
compute any missing wavefront propagations. To stress the algorithm, we reduced the
space available for the queue until we created two scenarios. In the first
scenario, the queue storage limit is exceeded once, and the wavefront
propagation execution phase has to be re-launched to fix any missing
propagations. In the second scenario, the queue storage limit is exceeded in
the second iteration of the wavefront propagation phase as well.  As a result,
a third iteration of this phase is needed to finish computation correctly.

The experimental results were obtained using the morphological reconstruction 
implementation and an single input tile of 4K$\times$4K pixels. The results 
show that performance penalty due to exceeding the queue storage limit is 
small. For instance, the performance hit is 6\% when the queue is exceeded once 
during the execution and 9\% when it is exceeded twice. In general, we observed 
that setting a storage limit that is 10\% larger than the initial queue size 
was sufficient. 
\subsection{Cooperative Multi-GPU Multi-CPU Execution}
These experiments evaluate the implementations when multiple GPUs and CPUs are
used cooperatively for execution. The input images have 100\% tissue coverage. 
The resolutions of the images are 96K$\times$96K pixels and 64K$\times$64K pixels
for the morphological reconstruction implementation and the distance transform
implementation, respectively; these are the maximum size images that can fit in
the main CPU memory for each algorithm. 

\begin{figure}[htb!]
\begin{center}
\mbox{
 \subfigure[Multi-GPU scalability.]{\label{fig:morph-multigpu} 
        \includegraphics[width=0.46\textwidth]{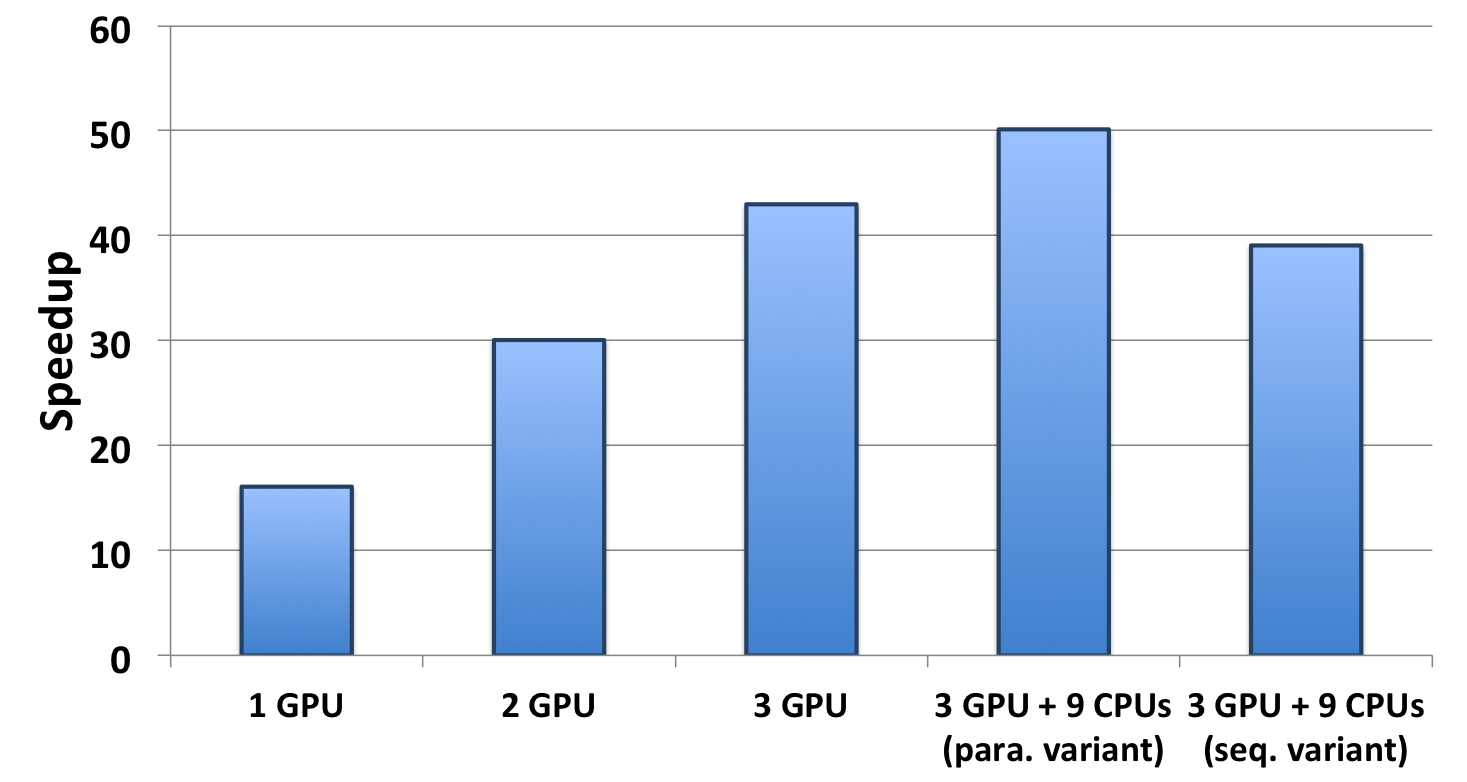}}
}
\mbox{
 \subfigure[Per Task Execution profile.]{\label{fig:morph-multigpu-profile} 
        \includegraphics[width=0.46\textwidth]{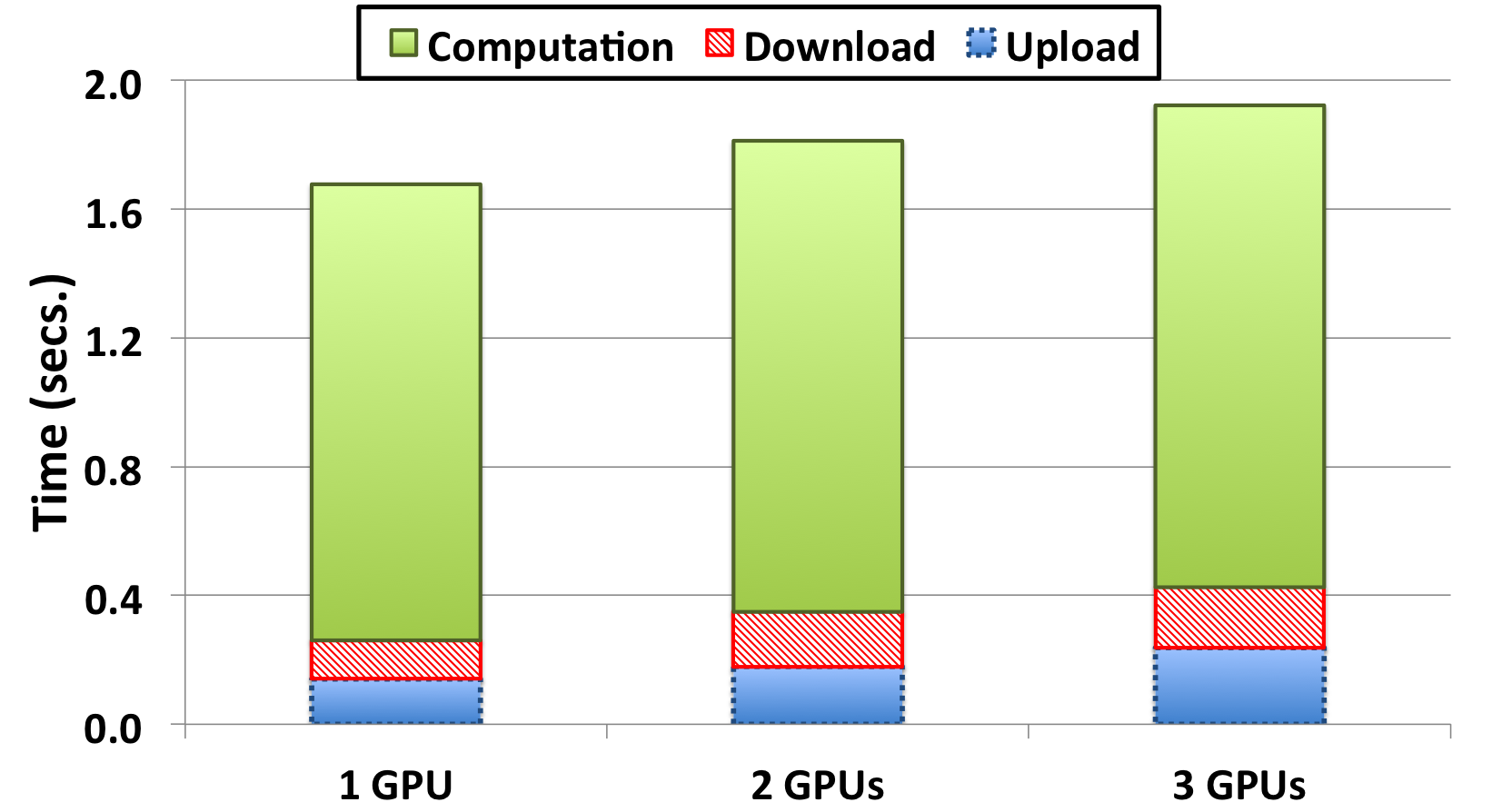}}
}
\caption{Performance of the morphological reconstruction implementation on multiple GPUs and using 
multiple CPUs and GPUs together.}
\label{fig:multigpu}
\end{center}
\end{figure}

The multi-GPU scalability of the morphological
reconstruction algorithm is presented in Figure~\ref{fig:morph-multigpu}. The
speedups achieved for one, two, and three GPUs, in comparison to the CPU single
core version, are about 16$\times$, 30$\times$ and 43$\times$. The
overall performance improvement is good, but the 3-GPUs execution achieves 
2.67$\times$ speedup with respect to the 1-GPU version which is below linear. 
To
investigate the reasons, we calculated the average cost of the wavefront propagation 
phase by breaking it down into three categories:
(i)~\emph{Computation}, which is the time spent by the application $kernels$;
(ii)~\emph{Download}, which corresponds to the time to download the results from GPU
to CPU memory; and (iii)~\emph{Upload}, which includes data copy from CPU to
GPU. The results, presented in Figure~\ref{fig:morph-multigpu-profile}, show
that there is an increase in data transfer costs as more GPUs are used,
limiting the scalability. We should note that we efficiently
utilized the architecture of the node, which is built with two I/O Hubs
connecting CPUs to GPUs (see Figure~\ref{fig:node}). In our approach, the CPU
thread managing a GPU is mapped to a CPU closer to the corresponding GPU. This 
allows for maximum performance to be attained during data transfers. When this
placement is not employed, the performance of the multi-GPU
scalability is degraded and only small performance improvements are observed 
when more than two GPUs are utilized.

Figure~\ref{fig:morph-multigpu} also presents the cooperative CPU--GPU
execution performance. In this scenario, when the multi-core version of 
the CPU operation is used (parallel variant), the multi-CPU multi-GPU executions achieve extra 
performance improvements of about 1.17$\times$ with respect to the 3-GPU 
configuration, or a total of 50$\times$ speedup on top of
the single core CPU execution. As is shown in the figure, 
the use of sequential CPU variants results in performance loss with respect to 
the multi-GPU only execution.  This performance degradation is the result of 
load imbalance among devices, which is higher with the sequential variants 
due to the larger gap in execution times between a GPU
execution and a single CPU core. The computation time for the 96K$\times$96K 
pixel image in the cooperative execution mode using 
all the CPUs and GPUs was 21 seconds.

\begin{figure}[htb!]
\begin{center}
	\includegraphics[width=0.6\textwidth]{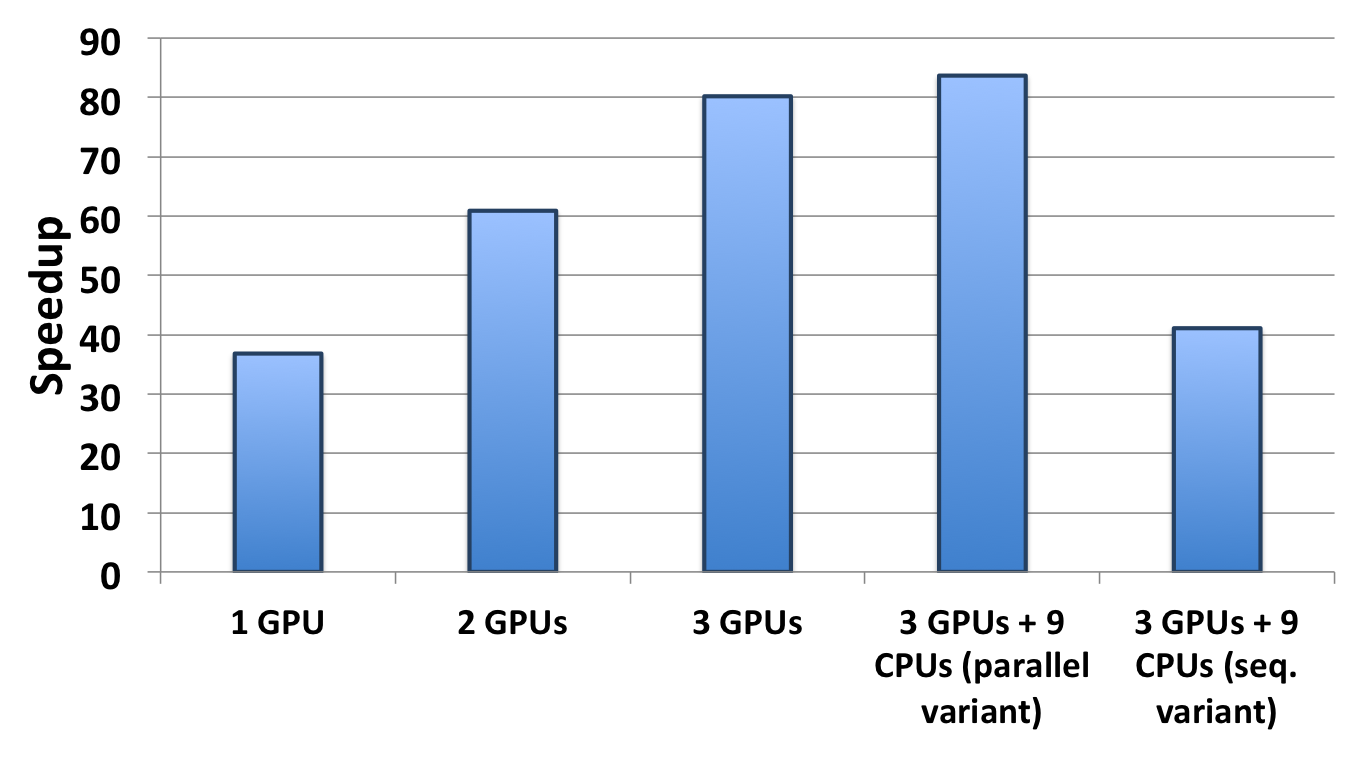}
\caption{Performance of the distance transform implementation on multiple 
GPUs and using multiple CPUs and GPUs together.}
\label{fig:dist-multigpu}
\end{center}
\end{figure}

Finally, Figure~\ref{fig:dist-multigpu} presents the cooperative multi-GPU/-CPU
execution results for distance transform. Similar to that of the morphological reconstruction 
implementation, the multi-GPU scalability
is good, but it is sub-linear because of increasing data transfer costs.
The cooperative CPU---GPU execution, in this case, resulted in a speedup of
1.07$\times$ on top of the execution with 3 GPUs. The smaller improvement
percentage, as compared to the morphological reconstruction algorithm, is
simply because of the better GPU acceleration achieved by distance transform.
The overall speedups of the algorithm when all GPUs and CPU cores are employed
together is about 85.6$\times$, as compared to the single CPU core execution.
For reference, the distance transform computation on the
64K$\times$64K pixel image took only 4.1 seconds.

\section{Related work} \label{sec:related}
The irregular wavefront propagation pattern has similarities to some 
fundamental graph scan algorithms, such as Breadth-First Search (BFS), but modified 
to support multiple sources. Much effort has been put into the development of parallel
implementations of BFS on multi-core CPUs~\cite{Agarwal:2010:SGE:1884643.1884670} and
GPUs~\cite{Hong:2011:ACG:1941553.1941590,Luo:2010:EGI:1837274.1837289}. The GPU
implementation proposed in~\cite{Luo:2010:EGI:1837274.1837289} is based on
a sequence of iterations until all nodes are visited. Each iteration visits all
nodes on the graph and processes those that are on the propagation frontier.
The work by Hong et al.~\cite{Hong:2011:ACG:1941553.1941590} targeted new
techniques to avoid load imbalance when nodes in a graph have a highly
irregular distribution of degrees (number of edges per node). In image operations 
with IWPP the approach proposed by Hong et al. would not do any better than
what was suggested by Karas~\cite{DBLP:conf/memics/Karas10}, since the degree of all
nodes (pixels) is the same and defined by the structuring element $G$. Moreover, 
the irregular wavefront propagation based approach visits those nodes (data elements) 
that are modified -- i.e., queued in the last iteration --, instead of all the nodes 
to compute only those that are in the frontier as 
in~\cite{Hong:2011:ACG:1941553.1941590}. 

The queue in our case is heavily employed and hundreds of thousands
to millions of nodes (pixels) may be queued per iteration. The solution
introduced in earlier work~\cite{Luo:2010:EGI:1837274.1837289}, if employed in
the IWPP, would require expensive inter-block communication at each iteration
of the application. This synchronization is realized by
synchronizing all thread blocks, stopping the $kernel$ execution, and
re-launching another round of computation. Unlike the previous
work~\cite{Luo:2010:EGI:1837274.1837289}, we propose that a scalable
hierarchical queue be employed in a per-block basis. This avoids 
communication among blocks and permits the entire computation to be efficiently
carried out into a single $kernel$ call. Moreover, since the connectivity of a 
pixel is known a priori, we can employ local queues more aggressively to
provide very fast local storage, as well as use parallel prefix-sum based
reductions to avoid expensive atomic operations when storing items into the
queue. Additionally, in our solution, an arbitrary large number of blocks may
be used, which enables the implementation to leverage GPU's dynamic block
scheduler in order to alleviate the effects of inter-block load imbalance.

Various image processing algorithms
have been ported to GPUs for efficient
execution~\cite{Fialka:2006:FCP:1153927.1154738,Korbes:2011:AWP:2023043.2023072,Scholl:2011:CMI:1938207.1938216}.
In addition, there is a growing set of libraries, such as the NVIDIA
Performance Primitives (NPP)~\cite{npp} and OpenCV~\cite{opencv_library}, which
encapsulate GPU-implementations of a number of image processing algorithms
through high level and flexible APIs. Most of the algorithms currently
available with this APIs, however, have more regular computation patterns as
compared to IWPP.

Morphological reconstruction has been increasingly employed by several
applications in the last decade. Efficient
algorithms~\cite{Vincent93morphologicalgrayscale,Laurent:1998:PIM:647051.759945}
and a number of applications were presented by
Vincent~\cite{Vincent:1991:ExEuDi}.  Other variants of Vincent's fast hybrid
algorithm (FH) have been proposed, such as the downhill filter
approach~\cite{Robinson:2004:EMR:1045937.1045948}. We have tested this approach
on our datasets and found it to be slower than FH. The high computational
demands of morphological reconstruction motivated other works that employed
specialized processors to speedup this algorithm. Jivet et
al.~\cite{Jivet:2008:ICE:1482247.1482253} implemented a version using FPGAs,
while Karas et al.~\cite{DBLP:conf/memics/Karas10} targeted GPUs. In both
cases, however, the parallel algorithms were designed on top of a sequential
baseline version that is intrinsically parallelizable, but about one order of
magnitude slower than the fastest sequential version. In our solution, on the
other hand, we extend the fastest baseline algorithm which exhibits the
irregular wavefront propagation pattern.  This approach has resulted in strong
and consistent performance improvements on top of the existing GPU
implementation by Karas~\cite{DBLP:conf/memics/Karas10}.

Euclidean distance transform is also important in image analysis and used in
several tasks as watershed based segmentation. This algorithm is very compute
demanding, which has attracted considerable attention in the last few years
regarding strategies for GPU
acceleration~\cite{Rong:2007:VJF:1270397.1271529,Rong:2006:JFG:1111411.1111431,DBLP:conf/visapp/SchneiderKW09,Cao:2010:PBA:1730804.1730818}.
Among the proposed solutions, the work of Schneider et
al.~\cite{DBLP:conf/visapp/SchneiderKW09} is the most similar to our distance
transform algorithm, because it also implements an output that is equivalent to
that of Danielsson's distance transform~\cite{danielsson80}. This GPU
implementation, however, is based on sweeps over the entire dataset, instead of
using propagations that process only those elements in the wavefront as in our
case. Moreover Schneider's algorithm only performs propagation of pixels in one
row at time, considering 2D problems, which limits GPU utilization.
Unfortunately, this code is not available for comparison, but the levels of
acceleration achieved in our case are higher than what Schneider's algorithm
attained. No multi-CPU multi-GPU algorithms have been developed for
morphological reconstruction and euclidean distance transform in the previous
works.

The use of CPUs and GPUs cooperatively has gained attention of the research
community in the last few years. Several works have developed
system and compiler techniques to support efficient execution on
heterogeneous CPU-GPU equipped
machines~\cite{merge,qilin09luk,cluster09george,1587427,hpdc10george,6061070,ravi2010compiler,6152715,hartley}.
In this paper, differently from previous work, we present a domain specific
framework for parallelization of irregular wavefront propagation based applications in
CPU-GPU equipped machines. This framework leverages the computation power of
such environments to a large class of applications by exporting a high level set
of tools, which may used by a programmer to accelerate applications fitting in the
irregular wavefront propagation pattern.

\section{Conclusions} \label{sec:conc}
The efficiency of the \emph{irregular wavefront propagation pattern} 
relies on tracking points in the wavefront and, consequently, avoiding 
unnecessary computation to
visit and process parts of the domain that do not contribute to the output.
This characteristic of the IWPP makes it an efficient computation structure 
in image analysis and other applications
~\cite{Vincent93morphologicalgrayscale,Vincent:1991:ExEuDi,
citeulike:557456,meyer90digital,lantuejoul80,Vincent92morphologicalarea,
Preparata:1985:CGI:4333,citeulike:3982757,Toussaint_1980}. Nevertheless, 
the parallelization of the IWPP on hybrid systems with 
multi-core CPUs and GPUs is complicated because of the dynamic, irregular, 
and data dependent data access and processing pattern. 
Our work has showed that a multi-level queue to track the 
wavefronts provides an efficient data structure for execution 
on a GPU. The multi-level queue structure enables the efficient 
use of fast memory hierarchies on a GPU. Inter- and intra-thread block
scalability can be achieved by per block queues
and parallel prefix-sum index calculation to avoid heavy use of atomic
operations.

We also have proposed a multi-processor execution strategy that 
divides the input domain into disjoint partitions (tiles) and assigns 
tiles to computing devices (CPU cores and GPUs) in a demand-driven 
fashion to reduce load imbalance. The tile based parallelization is
critical to achieving scalability on multi-core systems, since 
it reduces inter-thread interferences because of computation of 
the same regions of the domain without tiling. Inter-thread 
interferences cause inefficient use of cache. 

In order to evaluate the proposed approaches we have developed IWPP-based 
implementations of two widely used image processing operations, morphological
reconstruction and euclidean distance transform, on a state-of-the-art hybrid 
machine. In our experimental evaluation, these implementations have achieved 
orders of magnitude performance improvement (of upto 85$\times$) 
over the sequential CPU versions. Our results showed that coordinated use of 
CPUs and GPUs makes it feasible to process high resolution images in reasonable 
times, opening the way for large scale imaging studies. \\

\noindent {\bf Limitations of the Current Work.} This work focuses on the developing 
support for execution of the IWPP on shared memory hybrid machines equipped with 
multicore CPUs and GPUs. The current implementation does not support use of distributed
multi-node machines. The parallelization strategy we
employ for multiple GPUs and CPUs (Section~\ref{sec:multigpu}), however, can be
extended for the multi-node case. In this scenario, a messaging
passing mechanism should be employed to exchange information necessary for
propagations to cross partitions of the input domain assigned to different
machines. \\

\noindent {\bf Acknowledgments.} This research was funded, in part, by grants
from the National Institutes of Health through contract HHSN261200800001E by
the National Cancer Institute; and contracts 5R01LM009239-04 and
1R01LM011119-01 from the National Library of Medicine, R24HL085343 from the
National Heart Lung and Blood Institute, NIH NIBIB BISTI P20EB000591,
RC4MD005964 from National Institutes of Health, and PHS Grant UL1TR000454 from
the Clinical and Translational Science Award Program, National Institutes of
Health, National Center for Advancing Translational Sciences.  This research
used resources of the Keeneland Computing Facility at the Georgia Institute of
Technology, which is supported by the National Science Foundation under
Contract OCI-0910735. The content is solely the responsibility of the authors
and does not necessarily represent the official views of the NIH. We also want
to thank Pavel Karas for releasing the SR\_GPU implementation used in our
comparative evaluation.



%
\bibliographystyle{abbrv}
\bibliography{george}  
%

\end{document}